\documentclass[12pt]{article}
\newcommand{\be}{\begin{equation}}
\newcommand{\ee}{\end{equation}}
\newcommand{\bea}{\begin{eqnarray}}
\newcommand{\eea}{\end{eqnarray}}
\newcommand{\sptwo}{1.4}
\newcommand{\doublespace}{\edef\baselinestretch{\sptwo}\Large\normalsize}
\newcommand{\newsection}[1]{
\section{#1}
\setcounter{equation}{0}}

\renewcommand{\theequation}{\thesection.\arabic{equation}}
\newcounter{newapp}
\renewcommand{\thenewapp}{\Alph{newapp}}

\begin{document}
\begin{center}
{\large\bf Domain Walls, The Extended Superconformal Algebra \\
and \\
The Supercurrent}
\end{center}
\vspace*{0.15in}
\begin{center}
{\bf T.E. Clark}\footnote{e-mail address: clark@physics.purdue.edu}\\
{\it Department of Physics\\ 
Purdue University\\
West Lafayette, IN 47907-1396, U.S.A.}\\
~\\
{\bf Muneto Nitta}\footnote{e-mail address: nitta@phys-h.keio.ac.jp}\\
{\it Department of Physics\\ 
Keio University\\
Hiyoshi, Yokohama, Kanagawa 223-8521, Japan}\\
~\\
{\bf T. ter Veldhuis}\footnote{e-mail address: terveldhuis@macalester.edu}\\
{\it Department of Physics \& Astronomy\\ 
Macalester College\\
Saint Paul MN 55105-1899, U.S.A.}
~\\
\end{center}
~\\

\begin{center}
{\bf Abstract}
\end{center}

The presence of a domain wall is shown to require a tensorial central charge extension of the superconformal algebra.  The currents associated with the conformal central charges are constructed as spacetime moments of the SUSY tensorial central charge current.  The supercurrent is obtained and it contains the $R$ symmetry current, the SUSY spinor currents, the energy-momentum tensor and the SUSY tensorial central charge currents as its component currents.  All tensorial central charge extended superconformal currents are constructed from the supercurrent.  The superconformal currents' and the conformal tensorial central charge currents' (non-)conservation equations are expressed in terms of the generalized trace of the supercurrent.  It is argued that although the SUSY tensorial central charges are uncorrected, the conformal tensorial central charges receive radiative corrections.

\newpage
\doublespace


\newsection{Introduction}

The existence of point-like BPS particles in supersymmetric theories 
is known to require
a central extension of the  supersymmetry (SUSY) algebra.
Well known examples of such particles are kinks in $D=2$, which give rise to a central charge in the $~N=2$ SUSY algebra 
\cite{Witten:1978mh} and 
magnetic monopoles in $D=4$, which yield a central charge in the $~N=4 ~(N=2)$ SUSY algebra \cite{Osborn:1979tq}. 
At the same time,  the presence of extended objects, sometimes called $p$-branes, 
in supersymmetric theories 
implies a {\it tensorial} charge extension of the SUSY 
algebra \cite{deAzcarraga:1989gm}.
Since the additional tensorial charges  commute with the SUSY and translation charges, they are  referred to as central 
charges in the literature, even though they transform non-trivially under Lorentz transformations. 
D-branes or M-branes are the equivalent of such extended objects 
in string theory or M-theory; they
imply the central charge extension of the 
$D=10$ or $D=11$ SUSY algebra 
\cite{Townsend:1995gp}, \cite{Townsend:1997wg}.   
In field theory, topological solitons like domain walls,
vortex-strings, and junctions of walls and strings  play a role very similar to the branes of string theory.

The presence of a BPS domain wall in a $D=4, ~N=1$ supersymmetric field theory
results in a tensorial central charge which is in the $(1,0)$ representation of the Lorentz group
\cite{Chibisov:1997rc}, 
while the existence of a vortex-string or a domain-wall junction
requires a $(1/2,1/2)$ tensorial charge extension \cite{Gorsky:1999hk}. 
Possible tensorial charge extensions of  $N$ extended SUSY algebras in $D=3,4$ dimensions were
determined in \cite{Ferrara:1997tx}. 
Radiative corrections to tensorial central charges 
in the presence of various kinds of BPS topological solitons in 
supersymmetric field theory have also been studied
\cite{Shifman:1998zy}--\cite{Fujikawa:2003gi}, in particular in relation to soliton masses.

The minimal supersymmetric extension of the
$D=10$ conformal algebra (or, of the equivalent AdS$_{11}$  algebra)
was found to be $OSp(1|32)$ and necessitates the introduction of additional bosonic tensor charges \cite{Townsend:1997wz}. 
This algebra also appears as the tensorial central charge extension 
of superconformal algebras in lower dimensions with 
$32$ supercharges (see e.g. \cite{Craps:1999nc}, \cite{Kamimura:2003rx}).
Superconformal algebras with $N=1$ SUSY in various dimensions
were classified previously in \cite{vanHolten:1982mx}, including those with tensor charge extensions.
In the four dimensional case of interest here, 
the tensorial central charge extension of the
$D=4, ~N=1$ superconformal algebra  $SU(2,2|1)$
(or, of the equivalent $N=1$ SUSY AdS$_5$ algebra)
was found to be $OSp(1|8)$ \cite{Ferrara:1999si}.  It is a subalgebra of the $N=2,~ D=4$ centrally extended superconformal algebra presented in \cite{Bedding:1983uf}.

In light of  recent progress in 
$D=4$, $N=1$ superconformal field theories
in terms of a-maximization \cite{Intriligator:2003jj},
it seems timely to further discuss the tensorial charge extension of the
$D=4$, $N=1$ superconformal algebra. In this paper, the tensorial central
charges in the  superconformal algebra are studied, which arise due to the presence of a domain wall in $D=4, ~N=1$ supersymmetric field theories with chiral superfields.
The supercurrent is constructed and it is shown how all charges are associated with its components. As the considered theories are
not conformally invariant, the non-conservation of the conformal charges is analyzed.
The presence/absence of radiative corrections to 
the tensorial central charges is also discussed on general grounds. 

As noted above, the formation of a BPS saturated domain wall in $N=1, ~D=4$ superspace gives rise to a tensorial central charge extension of the SUSY algebra 
in which the Weyl spinor SUSY charges $Q_\alpha$ and $\bar{Q}_{\dot\alpha}$, with $\alpha,~\dot\alpha = 1,2$, self anticommute to yield the tensorial central charges $Z_{\alpha\beta}$ and $\bar{Z}_{\dot\alpha\dot\beta}$, respectively, 
\cite{Chibisov:1997rc} 
\be
\{Q_\alpha , Q_\beta \} = Z_{\alpha\beta} \qquad ; \qquad \{\bar{Q}_{\dot\alpha} , \bar{Q}_{\dot\beta} \} = \bar{Z}_{\dot\alpha\dot\beta}  .
\label{QQZ1}
\ee
The graded Jacobi identity applied to the angular momentum and the two $Q$ SUSY charges implies that $Z$ is in the (1,0) representation of the Lorentz group while applied to the angular momentum and the two $\bar{Q}$ SUSY charges implies that $\bar{Z}$ transforms as the (0,1) representation of the Lorentz group.  Likewise, the graded Jacobi identity applied to the $R$ charge and two $Q$ or two $\bar{Q}$ SUSY charges yields that the $R$-weights of the charges are additive, hence
\bea
[R, Z_{\alpha\beta} ] &=& +2 Z_{\alpha\beta} \cr
[R, \bar{Z}_{\dot\alpha\dot\beta} ] &=& -2 \bar{Z}_{\dot\alpha\dot\beta}.
\eea
This completes the tensorial central charge extension of the super-Poincar\'e algebra (see Appendix A for a comprehensive list of charges and commutators).

Within the context of a generalized Wess-Zumino model consisting of chiral fields $\phi^i$ and antichiral fields $\bar\phi^{\bar{i}}$, where $i, \bar{i}=1, 2, \ldots , N$, the action is given by
\be
\Gamma = \int dV ~K(\phi , \bar\phi) + \int dS ~W(\phi) + \int d\bar{S} ~\bar{W}(\bar\phi),
\label{WZModel1}
\ee
with (anti)chiral superpotential ($\bar{W}(\bar\phi)$) $W(\phi)$ and K\"ahler potential $K(\phi, \bar\phi)$.  Expanding the superfields in terms of their component fields
\bea
\phi^i &=& e^{-i\theta \rlap{/}{\partial} \bar\theta} \left[ A^i +\theta \psi^i +\theta^2 F^i \right]\cr
\bar\phi^{\bar{i}} &=& e^{+i\theta \rlap{/}{\partial} \bar\theta} \left[ \bar{A}^{\bar{i}} +\bar\theta \bar\psi^{\bar{i}} +\bar\theta^2 \bar{F}^{\bar{i}} \right],
\label{compfields}
\eea
Noether's theorem yields the form of the conserved SUSY currents 
\bea
Q_{N\alpha}^\mu &=& 16 K_{i\bar{i}} \left( \psi^i \sigma^\mu \bar\sigma^\lambda \partial_\lambda \bar{A}^{\bar{i}} \right)_\alpha -4i \bar{W}_{\bar{i}} \left( \sigma^\mu \bar\psi^{\bar{i}}\right)_\alpha \cr
 & & \cr
\bar{Q}_{N\dot\alpha}^\mu &=& 16 K_{i\bar{i}} \left( \partial_\lambda A^i \bar\sigma^\lambda \sigma^\mu \bar\psi^{\bar{i}}\right)_{\dot\alpha} +4i {W}_{{i}} \left( \psi^{{i}}\sigma^\mu \right)_{\dot\alpha} .
\label{Qcurrents1}
\eea
The operator conservation equations for the SUSY currents, $\partial_\mu Q^\mu_{N\alpha}(x)=0
$ and $\partial_\mu \bar{Q}^\mu_{N\dot\alpha}(x)=0$, follow from the SUSY invariance of the action (the Lagrangian varies into a total derivative) and the composite Euler-Lagrange equations which are assumed to vanish in the Born, that is tree, approximation.  (More specifically, operators are defined by their matrix elements which, via the LSZ reduction formalism, involve the time ordered functions of the fields with the inserted operator, in this case the Euler-Lagrange field equations times a field operator.  In turn the field equations cause the pole in the Green functions to be shifted from the mass shell and hence, when amputated and taken to shell, such terms vanish and the operator identity is obtained.) 

Calculating the SUSY variations of the SUSY Noether currents, or equivalently using the canonically quantized current algebra, yields the expression for the currents associated with the tensorial central charges \cite{Chibisov:1997rc} along with Euler-Lagrange composite field equation corrections, which again are taken to vanish on-shell,
\bea
\delta^Q_\alpha Q^\mu_{N\beta} + \delta^Q_\beta Q^\mu_{N\alpha} &=& Z^\mu_{N\alpha\beta} \cr
\delta^{\bar{Q}}_{\dot\alpha} \bar{Q}^\mu_{N\dot\beta} + \delta^{\bar{Q}}_{\dot\beta} \bar{Q}^\mu_{N\dot\alpha} &=& \bar{Z}^\mu_{N\dot\alpha\dot\beta}.
\label{QQZ2}
\eea
The topological form of the tensorial central charge Noether currents (the term central charge \lq\lq Noether" currents refers to the fact that the topological currents are derived from the SUSY variations of the SUSY Noether currents, being topological currents they are trivially conserved) are given by
\bea
Z^\mu_{N\alpha\beta} &=&-32i \sigma^{\mu\nu}_{\alpha\beta} \partial_\nu \bar{W} \cr
\bar{Z}^\mu_{N\dot\alpha\dot\beta} &=&-32i \bar\sigma^{\mu\nu}_{\dot\alpha\dot\beta} \partial_\nu {W}.
\label{topocurrent}
\eea
Off-shell the topological currents and the central charge currents will differ by Euler-Lagrange equations.  Since these are set to zero at present, the central charge currents have the above topological current form.  Integrating the zeroth component of the central charge current operators in equation (\ref{QQZ2}) reproduces the centrally extended algebra in equation (\ref{QQZ1}).  Further, from equation (\ref{topocurrent}), the central charges are given in terms of the superpotentials evaluated at spatial boundaries
\bea
Z_{\alpha\beta} &=& \int d^3 x Z^0_{N\alpha\beta} = -32 \sigma^{i}_{\alpha\beta}\int d^3 x \partial_{i} \bar{W} \cr
\bar{Z}_{\dot\alpha\dot\beta} &=& \int d^3 x \bar{Z}^0_{N\dot\alpha\dot\beta} = 32 \sigma^{i}_{\dot\alpha\dot\beta}\int d^3 x \partial_{i} {W} .
\eea
For a BPS domain wall in the $x$-$y$ plane, this yields non-zero central charges $16 \zeta \equiv Z_{22}=-Z_{11}$ and $16 \bar\zeta \equiv \bar{Z}_{11} = - \bar{Z}_{22}$ as the asymptotic difference of the superpotential evaluated at the different vacua connected by the domain wall which gives the domain wall tension
\bea
\zeta&=& +2 \int d^2 x \left( \bar{W}\Big\vert_{z=+\infty} -\bar{W}\Big\vert_{z=-\infty}\right) \cr
\bar{\zeta}&=& +2 \int d^2 x \left( {W}\Big\vert_{z=+\infty} -{W}\Big\vert_{z=-\infty}\right).
\label{topocharge}
\eea

The SUSY Noether currents can be improved so that they, $Q^\mu_\alpha$, $\bar{Q}^\mu_{\dot\alpha}$,  along with the improved Belinfante energy-momentum tensor, $T^{\mu\nu}$, the $R$ current, $R^\mu$, and the improved tensorial central charge currents, $Z^\mu_{\alpha\beta}$, $\bar{Z}^\mu_{\dot\alpha\dot\beta}$, are component fields in the superfield of currents known as the supercurrent \cite{Ferrara:1974pz}, \cite{Clark:1978jx}, \cite{PS1}, \cite{Clark:2004xv}.  It is found that the supercurrent, denoted $V^\mu = (\bar\sigma^\mu)^{\dot\alpha\alpha} V_{\alpha\dot\alpha}$, has the operator (on-shell) form
\bea
V^\mu (x, \theta, \bar\theta) &=& R^\mu (x) + i \theta\left[ Q^\mu (x) - \frac{1}{3}Q^\rho (x)\sigma_\rho \bar\sigma^\mu \right] -i \bar\theta\left[ \bar{Q}^\mu (x)-\frac{1}{3}\bar\sigma^\mu \sigma^\rho \bar{Q}_\rho (x)\right] \cr
 & & +\frac{1}{24}\theta^2 \sigma^{\mu\nu \beta}_\alpha Z_{\nu\beta}^{~~~\alpha} (x)-\frac{1}{24}\bar\theta^2 \bar\sigma^{\mu\nu \dot\alpha}_{~~~~~\dot\beta} \bar{Z}_{\nu~\dot\alpha}^{\dot\beta}(x)\cr
 & & +\theta \sigma_\nu \bar\theta \left\{ 2 \left[ T^{\mu\nu}(x) - \frac{1}{3} g^{\mu\nu} T^\rho_{~\rho}(x)\right] +\frac{1}{2}\epsilon^{\mu\nu\rho\sigma} \partial_\rho R_\sigma (x)\right\} \cr
 & & +\frac{1}{2}\theta^2 \bar\theta \left[ \partial^\mu \bar\sigma^\nu -i\epsilon^{\mu\nu\rho\sigma} \bar\sigma_\rho \partial_\sigma \right] \left[ Q^\mu (x)- \frac{1}{3}Q^\rho (x)\sigma_\rho \bar\sigma^\mu \right] \cr
 & & -\frac{1}{2} \bar\theta^2 \theta  \left[ \partial^\mu \sigma^\nu -i\epsilon^{\mu\nu\rho\sigma} \sigma_\rho \partial_\sigma \right]\left[ \bar{Q}^\mu (x)-\frac{1}{3}\bar\sigma^\mu \sigma^\rho \bar{Q}_\rho (x)\right]\cr
 & & +\frac{1}{4}\theta^2 \bar\theta^2 \left[ \partial^2 g^{\mu\nu} -2 \partial^\mu \partial^\nu \right] R_\nu (x).
\label{supercurrentcomp}
\eea
Hence all of the tensorial central charge extended super-Poincar\'e currents can be obtained from the supercurrent \cite{Clark:1978jx}.  Indeed the currents are the $\theta$ and $\bar\theta$ independent component of the superfield expressions (denoted by hats) obtained from $V^\mu$ by differentiation
\bea
\hat{R}^\mu &=& V^\mu = R^\mu + \cdots \cr
\hat{Q}^\mu_\alpha &=& -i\left[ D_\alpha V^\mu -(DV^\nu \sigma_\nu \bar\sigma^\mu)_\alpha \right] = Q^\mu_\alpha + \cdots \cr
\hat{\bar{Q}}^\mu_{\dot\alpha} &=& +i\left[ \bar{D}_{\dot\alpha} V^\mu -( \bar\sigma_\mu \sigma^\nu \bar{D} V_\nu)_{\dot\alpha} \right] = \bar{Q}^\mu_{\dot\alpha} + \cdots \cr
\hat{Z}^\mu_{\alpha\beta} &=& \sigma^{\mu\nu}_{\alpha\beta} DD V_\nu =+2i\left[ \bar\sigma^{\mu\gamma}_\alpha D_\beta +\bar\sigma^{\mu\gamma}_\beta D_\alpha \right] D^\gamma 
V_{\gamma\dot\gamma} =Z^\mu_{\alpha\beta} +\cdots \cr
\hat{\bar{Z}}^\mu_{\dot\alpha\dot\beta} &=& -\bar\sigma^{\mu\nu}_{\dot\alpha\dot\beta} \bar{D}\bar{D} V_\nu =-2i\left[ \bar{D}_{\dot\alpha} \bar\sigma^{\mu\dot\gamma}_{\dot\beta}  +\bar{D}_{\dot\beta} \bar\sigma^{\mu\dot\gamma}_{\dot\alpha}\right] \bar{D}^{\dot\gamma} 
V_{\gamma\dot\gamma} = \bar{Z}^\mu_{\dot\alpha\dot\beta} +\cdots \cr
\hat{T}^{\mu\nu} &=& +\frac{1}{16}\left[ V^{\mu\nu} +V^{\nu\mu} - 2g^{\mu\nu} V^\rho_{~\rho} \right] = T^{\mu\nu} +\cdots ,
\label{currents}
\eea
where $V^{\mu\nu} \equiv (D\sigma^\mu \bar{D} - \bar{D} \bar\sigma^\mu D) V^\nu$ and the ellipses denote the higher $\theta$ and $\bar\theta$ components of the superfields.  The angular momentum tensor is made from the energy-momentum tensor as usual, $\hat{M}^{\mu\nu\rho}= x^\rho \hat{T}^{\mu\nu} -x^\nu \hat{T}^{\mu\rho} = M^{\mu\nu\rho} +\cdots$.

The existence of the tensorial central charges $Z_{\alpha\beta}$ and $\bar{Z}_{\dot\alpha\dot\beta}$
requires that the superconformal algebra be additionally central charge extended.  In particular the special (or conformal) SUSY charges $S_\alpha$ and $\bar{S}_{\dot\alpha}$ self anticommute to give the nonzero tensorial central charges $X_{\alpha\beta}$ and $\bar{X}_{\dot\alpha\dot\beta}$
\be
\{S_\alpha , S_\beta \} = X_{\alpha\beta} \qquad ; \qquad \{\bar{S}_{\dot\alpha} , \bar{S}_{\dot\beta} \} = \bar{X}_{\dot\alpha\dot\beta}  .
\label{SSX1}
\ee
Also the SUSY and conformal SUSY charges anticommute to yield the vectorial central charges $Y^\mu$ and $\bar{Y}^\mu$
\be
\{Q_\alpha , \bar{S}_{\dot\alpha} \} = Y_{\alpha\dot\alpha}=+2\sigma^\mu_{\alpha\dot\alpha} Y_\mu \qquad ; \qquad \{{S}_{\alpha} , \bar{Q}_{\dot\alpha} \} = \bar{Y}_{\alpha\dot\alpha} = +2\sigma^\mu_{\alpha\dot\alpha} \bar{Y}_\mu .
\label{QSY1}
\ee
The complete extended superconformal algebra is given in Appendix A.  The purpose of this paper is to construct these tensorial central charges and their currents by use of the associated current algebra.  It is shown that they are given by space-time moments of the super-Poincar\'e (SUSY) tensorial central charge currents $Z^\mu_{\alpha\beta}$ and $\bar{Z}^\mu_{\dot\alpha\dot\beta}$ similar to the way the superconformal charges and currents are moments of the SUSY currents and the energy-momentum tensor.  The generalized trace of the supercurrent given by $D^\alpha V_{\alpha\dot\alpha}$ and $\bar{D}^{\dot\alpha} V_{\alpha\dot\alpha}$ determines all conservation and non-conservation equations for the centrally extended superconformal currents.  

In section 2 the current algebra is used to determine the central charges and their currents in the framework of the generalized Wess-Zumino model of equation (\ref{WZModel1}).  The conformal tensorial central charge currents and their nonconservation equations follow from that of the SUSY tensorial central charge currents, $Z^\mu_{\alpha\beta}$ and $\bar{Z}^\mu_{\dot\alpha\dot\beta}$, and their traces.  This in turn is given by the generalized trace of the supercurrent, as are all the superconformal current nonconservation equations.  In section 3 the component currents of the supercurrent are determined.  The $R$ symmetry current, the modified and improved SUSY currents and energy-momentum tensor and SUSY tensorial central charge currents are explicitly determined in terms of the component fields.  Radiative corrections to the SUSY tensorial central charges and currents have been extensively studied in models containing extended objects in various dimensions and types of SUSY \cite{Shifman:1998zy}, \cite{Chibisov:1997rc}, \cite{Shizuya:2005th}-\cite{Fujikawa:2003gi}.  Although a rigorous treatment of the radiative corrections to the central charge currents is beyond the scope of this paper, the form of the radiative corrections to the supercurrent trace equations is conjectured upon in section 4.  There it is argued, that while the SUSY tensorial central charges evaluated on a domain wall are not radiatively corrected, the conformal tensorial central charges depend on the anomalous dimension of the superfields.

\newsection{Tensorial Central Charges and Currents}

The SUSY and superconformal variations of the associated currents yield expressions for the various tensorial central charges and currents.  The superconformal currents, denoted $S^\mu_{N\alpha}$ and $\bar{S}^\mu_{N\dot\alpha}$, corresponding to the action in equation (\ref{WZModel1}), are found via Noether's theorem using the superconformal transformations given in Appendix B
\bea
S^\mu_{N\alpha} &=& - {\rlap{/}{x}}_{\alpha\dot\alpha}~\bar{Q}^{\mu\dot\alpha}_{N} + 2 s^\mu_\alpha\cr
\bar{S}^\mu_{N\dot\alpha} &=& - {Q}^{\mu\alpha}_{N}~{\rlap{/}{x}}_{\alpha\dot\alpha}  + 2 \bar{s}^\mu_{\dot\alpha} ,
\eea
where the $s^\mu_\alpha$ and $\bar{s}^\mu_{\dot\alpha}$ terms are given by
\bea
s^\mu_\alpha &=& -16 d A^i K_{i\bar{i}} (\bar\sigma^\mu \bar\psi^{\bar{i}})_\alpha \cr
\bar{s}^\mu_{\dot\alpha} &=& -16 d (\psi^{i}\sigma^\mu )K_{i\bar{i}} \bar{A}^{\bar{i}} _\alpha ,
\eea
with $d$ the scale weight of the chiral fields which, canonically, is given by $d=1$ and will receive anomalous dimension radiative corrections such that $d\rightarrow 1+\gamma$  as obtained in section 4.  The action is not superconformal invariant so that the divergence of the superconformal current yields
\bea
\partial_\mu S^\mu_{N\alpha} &=& \Delta^{S}_{\alpha} {\cal L} \cr
\partial_\mu \bar{S}^\mu_{N\dot\alpha} &=& \Delta^{\bar{S}}_{\dot\alpha} {\cal L}  ,
\eea
with the superconformal symmetry breaking terms given by
\bea
\Delta^{S}_{\alpha} {\cal L} &=& +8i \psi^{i}_{\alpha} \frac{\partial}{\partial {A}^{i}} \left[ 3 -d {{\cal N}} \right] {W} - 16i \bar{F}^{\bar{i}} \psi^{{i}}_{\alpha} \left[ 2 -2d -d {{\cal N}} \right]K_{i\bar{i}} \cr
 & & +4i \bar\psi^{\bar{i}} \bar\psi^{\bar{j}} \psi^{{i}}_{\alpha} \left[ 2 -2d -d {{\cal N}} \right]K_{i\bar{i}\bar{j}} +32 (\sigma^\mu \bar\psi^{\bar{i}} )_{\alpha} \partial_\mu {A}^{{i}} \left[ 1 -1d -d \bar{{\cal N}} \right]K_{i\bar{i}} \cr
 & & \cr
\Delta^{\bar{S}}_{\dot\alpha} {\cal L} &=& -8i \bar\psi^{\bar{i}}_{\dot\alpha} \frac{\partial}{\partial \bar{A}^{\bar{i}}} \left[ 3 -d \bar{{\cal N}} \right] \bar{W} + 16i F^i \bar\psi^{\bar{i}}_{\dot\alpha} \left[ 2 -2d -d \bar{{\cal N}} \right]K_{i\bar{i}} \cr
 & & -4i \psi^i \psi^j \bar\psi^{\bar{i}}_{\dot\alpha} \left[ 2 -2d -d \bar{{\cal N}} \right]K_{ij\bar{i}} +32 (\psi^i \sigma^\mu)_{\dot\alpha} \partial_\mu \bar{A}^{\bar{i}} \left[ 1 -1d -d \bar{{\cal N}} \right]K_{i\bar{i}} .\cr
 & & 
\eea
The number operators are given by the functional operators ${\cal N} = \int d^4 x A^i (x) \delta/\delta A^i(x)$ and $\bar{{\cal N}} = \int d^4 x \bar{A}^{\bar{i}}(x) \delta/\delta \bar{A}^{\bar{i}}(x)$.  

From the form of the extended superconformal algebra, equation (\ref{QSY1}), the additional central charge currents can be found by considering the SUSY variation of the superconformal current plus the superconformal variation of the SUSY current
\bea
Y^\mu_{N\alpha\dot\alpha} \equiv \delta^Q_\alpha \bar{S}^\mu_{N\dot\alpha} + \delta^{\bar{S}}_{\dot\alpha} Q^\mu_{N\alpha} &=& \rlap{/}{x}^{\beta}_{~\dot\alpha} \left[ \delta^Q_\alpha Q^\mu_{N\beta} + \delta^Q_{\beta} Q^\mu_{N\alpha}\right]-32 d \sigma^\mu_{\alpha\dot\alpha} \bar{W,}_{\bar{i}} \bar{A}^{\bar{i}} \cr
 & & \cr
 &=& -32i (\bar{\rlap{/}{x}}\sigma^{\mu\nu})_{\dot\alpha\alpha} \partial_\nu \bar{W} -32 d \sigma^\mu_{\alpha\dot\alpha} \bar{W,}_{\bar{i}} \bar{A}^{\bar{i}} ,
\eea
where once again Euler-Lagrange terms have been set to zero.  Likewise the conjugate central charge current is given by
\bea
\bar{Y}^\mu_{N\alpha\dot\alpha} \equiv \delta^{\bar{Q}}_{\dot\alpha} {S}^\mu_{N\alpha} + \delta^{{S}}_{\alpha} \bar{Q}^\mu_{N\dot\alpha} &=& -\rlap{/}{\bar{x}}^{\dot\beta}_{~\alpha} \left[ \delta^{Q}_{\dot\alpha} \bar{Q}^\mu_{N\dot\beta} + \delta^{\bar{Q}}_{\dot\beta} \bar{Q}^\mu_{N\dot\alpha}\right]-32 d \sigma^\mu_{\alpha\dot\alpha} {W,}_{{i}} {A}^{{i}} \cr
 & & \cr
 &=& -32i (\rlap{/}{x}\bar\sigma^{\mu\nu})_{\alpha\dot\alpha} \partial_\nu {W} -32 d \sigma^\mu_{\alpha\dot\alpha} {W,}_{{i}} {A}^{{i}} ,
\eea
Also the superconformal variations of the superconformal currents, equation (\ref{SSX1}), yield
\be
\bar{X}^\mu_{N\dot\alpha\dot\beta}\equiv\delta^{\bar{S}}_{\dot\alpha} \bar{S}^\mu_{N\dot\beta} + \delta^{\bar{S}}_{\dot\beta} \bar{S}^\mu_{N\dot\alpha} = 32i (\bar{\rlap{/}{x}}\sigma^{\mu\nu}\rlap{/}{x})_{\dot\alpha\dot\beta} \partial_\nu \bar{W}
-64i d x_\nu \bar{\sigma}^{\mu\nu}_{\dot\alpha\dot\beta} \bar{A}^{\bar{i}}\bar{W,}_{\bar{i}} 
\ee
and likewise for the conjugate current
\be
{X}^\mu_{N\alpha\beta}\equiv\delta^{{S}}_{\alpha} {S}^\mu_{N\beta} + \delta^{{S}}_{\beta} {S}^\mu_{N\alpha} = 32i ({\rlap{/}{x}}\bar\sigma^{\mu\nu}\rlap{/}{\bar{x}})_{\alpha\beta} \partial_\nu {W}
-64i d x_\nu {\sigma}^{\mu\nu}_{\alpha\beta} {A}^{{i}}{W,}_{{i}} .
\ee

Belinfante improved SUSY and superconformal currents can be defined.  The improved SUSY currents, denoted $Q^\mu_{\alpha}$ and $\bar{Q}^\mu_{\dot\alpha}$, are given by
\bea
Q^\mu_{\alpha} &=& Q^\mu_{N\alpha} + \partial_\rho G^{\rho\mu}_{\alpha} \cr
\bar{Q}^\mu_{\dot\alpha} &=& \bar{Q}^\mu_{N\dot\alpha} + \partial_\rho \bar{G}^{\rho\mu}_{\dot\alpha} ,
\eea
with the antisymmetric tensor indexed improvement terms 
\bea
G^{\rho\mu}_{\alpha} &=& a K_{i\bar{i}} ( \psi^i \sigma^{\rho\mu})_\alpha \bar{A}^{\bar{i}} \cr
\bar{G}^{\rho\mu}_{\dot\alpha} &=& \bar{a} K_{i\bar{i}}{A}^{{i}} ( \bar\sigma^{\rho\mu}\bar\psi^{\bar{i}} )_{\dot\alpha}.
\eea
The arbitrary constants $a$ and $\bar{a}$ can be chosen so that the improved superconformal currents, denoted $S^\mu_\alpha$ and $\bar{S}^\mu_{\dot\alpha}$, are simply the space-time moment of the improved SUSY currents
\bea
S^\mu_{\alpha} &=& S^\mu_{N\alpha} +\partial_\rho [ \rlap{/}{x}_\alpha^{~\dot\alpha} \bar{G}^{\rho\mu}_{\dot\alpha}] =- \rlap{/}{x}_{\alpha\dot\alpha}\bar{Q}^{\mu\dot\alpha} + \left(2 -\frac{3i\bar{a}}{16d}\right){s}^\mu_{\alpha}   \cr
 & & \cr
\bar{S}^\mu_{\dot\alpha} &=& \bar{S}^\mu_{N\dot\alpha} +\partial_\rho [ \bar{\rlap{/}{x}}^{~\alpha}_{\dot\alpha} {G}^{\rho\mu}_{\alpha} ]
=-Q^{\mu\alpha} \rlap{/}{x}_{\alpha\dot\alpha} + \left(2 -\frac{3ia}{16d}\right)\bar{s}^\mu_{\dot\alpha} .
\label{impS}
\eea
Choosing $a=-i\frac{32d}{3}$ and $\bar{a}=+i \frac{32d}{3}$, the desired forms of the superconformal currents are obtained
\bea
S^\mu_{\alpha} &=&- \rlap{/}{x}_{\alpha\dot\alpha}\bar{Q}^{\mu\dot\alpha}\cr
\bar{S}^\mu_{\dot\alpha} &=&-Q^{\mu\alpha} \rlap{/}{x}_{\alpha\dot\alpha}.
\label{superconformalmoment}
\eea

Applying the SUSY and superconformal variations to these improved currents, the improved current algebra and the improved central charge currents are obtained
\bea
\delta^Q_\alpha Q^\mu_{\beta} + \delta^Q_\beta Q^\mu_{\alpha} &=& Z^\mu_{\alpha\beta} \qquad ;\qquad
\delta^{\bar{Q}}_{\dot\alpha} \bar{Q}^\mu_{\dot\beta} + \delta^{\bar{Q}}_{\dot\beta} \bar{Q}^\mu_{\dot\alpha} = \bar{Z}^\mu_{\dot\alpha\dot\beta}\cr
\delta^Q_\alpha \bar{S}^\mu_{\dot\alpha} + \delta^{\bar{S}}_{\dot\alpha} Q^\mu_{\alpha} &=&  Y^\mu_{\alpha\dot\alpha}
\qquad ;\qquad
\delta^{\bar{Q}}_{\dot\alpha} {S}^\mu_{\alpha} + \delta^{{S}}_{\alpha} \bar{Q}^\mu_{\dot\alpha} =  \bar{Y}^\mu_{\alpha\dot\alpha}\cr
\delta^{{S}}_{\alpha} {S}^\mu_{\beta} + \delta^{{S}}_{\beta} {S}^\mu_{\alpha} &=& {X}^\mu_{\alpha\beta}
\qquad ;\qquad
\delta^{\bar{S}}_{\dot\alpha} \bar{S}^\mu_{\dot\beta} + \delta^{\bar{S}}_{\dot\beta} \bar{S}^\mu_{\dot\alpha} = \bar{X}^\mu_{\dot\alpha\dot\beta} ~~,
\eea
where the improved central charge currents are found to be (ignoring Euler-Lagrange terms)
\bea
Z^\mu_{\alpha\beta} &=& -32i \sigma^{\mu\nu}_{\alpha\beta} \partial_\nu \left[ \bar{W} -\frac{d}{3} \bar{A}^{\bar{i}} \bar{W,}_{\bar{i}}\right] 
\quad ;\quad
\bar{Z}^\mu_{\dot\alpha\dot\beta} = -32i \bar\sigma^{\mu\nu}_{\dot\alpha\dot\beta} \partial_\nu \left[ {W} -\frac{d}{3} {A}^{{i}} {W,}_{{i}} \right]\cr
Y^\mu_{\alpha\dot\alpha} &=& (\bar{\rlap{/}{x}}Z^\mu)_{\dot\alpha\alpha} 
\qquad\qquad\qquad\qquad\quad ;\quad
\bar{Y}^\mu_{\alpha\dot\alpha} = ({\rlap{/}{x}}\bar{Z}^\mu)_{\alpha\dot\alpha} \cr
X^\mu_{\alpha\beta}&=& -(\rlap{/}{x}\bar{Z}^\mu \bar{\rlap{/}{x}})_{\alpha\beta}
\qquad\qquad\qquad\quad\quad ;\quad
\bar{X}^\mu_{\dot\alpha\dot\beta}= -(\bar{\rlap{/}{x}}Z^\mu \rlap{/}{x})_{\dot\alpha\dot\beta} ~~.
\label{tencentcurrents}
\eea
The conformal central charge currents $Y^\mu_{\alpha\dot\alpha}$, $\bar{Y}^\mu_{\alpha\dot\alpha}$, $X^\mu_{\alpha\beta}$ and $\bar{X}^\mu_{\dot\alpha\dot\beta}$ are given by the space-time moments of the topological SUSY central charges $Z^\mu_{\alpha\beta}$ and $\bar{Z}^\mu_{\dot\alpha\dot\beta}$.  The topological central charge currents are trivially conserved, however the conformal central charge currents are not conserved
\bea
\partial_\mu Y^\mu_{\alpha\dot\alpha} &=& (\bar\sigma_\mu Z^\mu )_{\alpha\dot\alpha} = 96 \sigma^\mu_{\alpha\dot\alpha} \partial_\mu \left[ \bar{W} -\frac{d}{3} \bar{A}^{\bar{i}} \bar{W,}_{\bar{i}}\right] \cr
\partial_\mu \bar{Y}^\mu_{\alpha\dot\alpha} &=& (\sigma_\mu \bar{Z}^\mu )_{\alpha\dot\alpha} = 96  \sigma^\mu_{\alpha\dot\alpha} \partial_\mu \left[ {W} -\frac{d}{3} {A}^{{i}} {W,}_{{i}}\right] \cr
\partial_\mu X^\mu_{\alpha\beta} &=& -(\sigma_\mu \bar{Z}^\mu \bar{\rlap{/}{x}})_{\alpha\beta}-(\sigma_\mu \bar{Z}^\mu \bar{\rlap{/}{x}})_{\beta\alpha} = -96i\sigma^{\mu\nu}_{\alpha\beta}[x_\mu \partial_\nu -x_\nu \partial_\mu ]\left[ {W} -\frac{d}{3} {A}^{{i}} {W,}_{{i}}\right] \cr
\partial_\mu \bar{X}^\mu_{\dot\alpha\dot\beta} &=&   -(\bar{\rlap{/}{x}} {Z}^\mu \sigma_\mu)_{\dot\alpha\dot\beta}  -(\bar{\rlap{/}{x}} {Z}^\mu \sigma_\mu)_{\dot\beta\dot\alpha} = -96i \bar\sigma^{\mu\nu}_{\dot\alpha\dot\beta}[x_\mu \partial_\nu -x_\nu \partial_\mu ]\left[ \bar{W} -\frac{d}{3} \bar{A}^{\bar{i}} \bar{W,}_{\bar{i}}\right]. \cr
 & & 
\label{centralcurrentsdivergence}
\eea
The nonconservation equations of the conformal central charge currents are determined by the vector component of the trace of the topological SUSY central charge currents, $(\sigma_\mu \bar{Z}^\mu)_{\alpha\dot\alpha}$ and $(Z^\mu \sigma_\mu )_{\alpha\dot\alpha}$.  Finally, the $Q$-$S$ anticommutators give rise to the angular momentum, dilatation and $R$ currents as seen in equation (\ref{rhofactor}).  Since the domain wall solution is a static solution its $R$ charge vanishes, hence the $\rho=-1/3$ factor in the algebra cannot be tested.

\newsection{The Supercurrent}

Not only are the tensorial central charge extended super-Poincar\'e symmetry currents and their conservation equations obtained from the supercurrent and its derivatives as indicated in equation (\ref{currents}), but also the centrally extended superconformal currents are obtained from the supercurrent.  For instance the superconformal symmetry current is given by  
\bea
\hat{S}^\mu_\alpha &=& -\rlap{/}{x}_{\alpha\dot\alpha} \hat{\bar{Q}}^{\mu\dot\alpha} = -i\rlap{/}{x}_{\alpha\dot\alpha} \left[ \bar{D}^{\dot\alpha} V^\mu -( \bar\sigma^\mu \sigma_\nu \bar{D} V^\nu)^{\dot\alpha} \right] = {S}^\mu_{\alpha} + \cdots \cr
\hat{\bar{S}}^\mu_{\dot\alpha} &=& -\hat{{Q}}^{\mu\alpha}\rlap{/}{x}_{\alpha\dot\alpha}  = +i \left[ 
D^\alpha V^\mu -(DV^\nu \sigma_\nu \bar\sigma^\mu)^\alpha \right] \rlap{/}{x}_{\alpha\dot\alpha}= \bar{S}^\mu_{\dot\alpha} + \cdots .
\label{Scurrents}
\eea
The divergence of the superconformal current is found from above to be 
\bea
\partial_\mu \hat{S}^\mu_\alpha &=& -\sigma_{\mu\alpha\dot\alpha} \hat{\bar{Q}}^{\mu\dot\alpha} = +6i\bar{D}^{\dot\alpha} V_{\alpha\dot\alpha} \cr
\partial_\mu \hat{\bar{S}}^\mu_{\dot\alpha} &=& -\hat{{Q}}^{\mu\alpha}\sigma_{\mu\alpha\dot\alpha} = -6i{D}^{\alpha} V_{\alpha\dot\alpha}.
\eea
The nonconservation of the superconformal currents are characterized by the generalized traces of the supercurrent, ${D}^{\alpha} V_{\alpha\dot\alpha}$ and $\bar{D}^{\dot\alpha} V_{\alpha\dot\alpha}$.  
Likewise the nonconservation equations of the superconformal tensorial central charge currents are given in equation (\ref{centralcurrentsdivergence}) by the traces of the tensorial SUSY central charge currents $(\sigma_\mu \bar{Z}^\mu)_{\alpha\dot\alpha}$ and $(Z^\mu \sigma_\mu )_{\alpha\dot\alpha}$.  In turn, from equation (\ref{currents}), it is seen that the sigma traces of $Z$ and $\bar{Z}$ are given in terms of the generalized traces of the supercurrent
\bea
(\hat{Z}^\mu \sigma_\mu)_{\beta\dot\alpha} &=& +12i D_\beta D^\alpha V_{\alpha\dot\alpha}\cr
(\sigma_\mu \hat{\bar{Z}}^\mu)_{\alpha\dot\beta} &=& -12i \bar{D}_{\dot\beta} \bar{D}^{\dot\alpha} V_{\alpha\dot\alpha} .
\eea
Indeed the SUSY tensorial central charge currents are themselves given in terms of the generalized traces of the supercurrent as is already evident in equation (\ref{currents}).

The supercurrent associated with the generalized Wess-Zumino action, equation (\ref{WZModel1}), is
given by
\bea
V_{\alpha\dot\alpha}&=& -\frac{8}{3}D_\alpha \phi^i K,_{i\bar{i}} \bar{D}_{\dot\alpha} \bar\phi^{\bar{i}} -\frac{16}{3}i K,_{i} \rlap{/}{\partial}_{\alpha\dot\alpha} \phi^i +\frac{16}{3}i K,_{\bar{i}} \rlap{/}{\partial}_{\alpha\dot\alpha} \bar\phi^{\bar{i}} \cr
 &=& -\frac{4}{3}[D_\alpha , \bar{D}_{\dot\alpha}]K -\frac{8}{3}iK,_{i}\rlap{/}{\partial}_{\alpha\dot\alpha} \phi^i +\frac{8}{3}i K,_{\bar{i}} \rlap{/}{\partial}_{\alpha\dot\alpha}  \bar\phi^{\bar{i}} .
\eea
Using the equations of motion as algebraic identities in which the Euler-Lagrange derivatives are not set to zero but are kept explicit so that
\bea
\frac{\delta \Gamma}{\delta \phi^i} &=& \bar{D}\bar{D} K,_{i} + W,_{i} \cr
\frac{\delta \Gamma}{\delta \bar\phi^{\bar{i}}} &=& {D}{D} K,_{\bar{i}} + \bar{W},_{\bar{i}},
\label{fieldequations}
\eea
the generalized traces of the supercurrent take the form
\bea
-2w_\alpha \Gamma &=& \bar{D}^{\dot\alpha} V_{\alpha\dot\alpha} +2D_\alpha S \cr
-2\bar{w}_{\dot\alpha} \Gamma &=& {D}^{\alpha} V_{\alpha\dot\alpha} +2\bar{D}_{\dot\alpha} \bar{S} .
\label{supercurrenttrace}
\eea
The Ward-Takahashi identity functional differential operators, $w_\alpha$ and $\bar{w}_{\dot\alpha}$, acting on the action functional $\Gamma$ are defined by
\bea
w_\alpha \Gamma &=& 2(D_\alpha \phi^i ) \frac{\delta \Gamma}{\delta \phi^i} + \sum_i n_i D_\alpha\left(\phi^i \frac{\delta \Gamma}{\delta \phi^i} \right)\cr
\bar{w}_{\dot\alpha} \Gamma &=& 2(\bar{D}_{\dot\alpha} \bar\phi^{\bar{i}} ) \frac{\delta \Gamma}{\delta \bar\phi^{\bar{i}}} + \sum_{\bar{i}} \bar{n}_{\bar{i}} \bar{D}_{\dot\alpha}\left(\bar\phi^{\bar{i}} \frac{\delta \Gamma}{\delta \bar\phi^{\bar{i}}} \right).
\eea
The superconformal algebra closes on the chiral fields only if the $R$-weights, $n_i$ and $\bar{n}_{\bar{i}}$, are related to the scaling weights, $d_i$ and $\bar{d}_{\bar{i}}$, of the fields ($\bar{d}_i =d_i$):
\bea
n_i &=& -\frac{2}{3}d_i \cr
\bar{n}_{\bar{i}} &=& +\frac{2}{3} \bar{d}_{\bar{i}} .
\eea
The potential symmetry breaking terms in the traces are given by the chiral $S$ and antichiral $\bar{S}$ fields
\bea
S&=& -\left[ 2W +\sum_i n_i \phi^i W,_{i} + \bar{D}\bar{D} ( \frac{2}{3}K +\sum_i n_i \phi^i K,_{i} )\right] \cr
\bar{S}&=& -\left[ 2\bar{W} +\sum_{\bar{i}} \bar{n}_{\bar{i}} \bar\phi^{\bar{i}} \bar{W},_{\bar{i}} + {D}{D} ( \frac{2}{3}K +\sum_{\bar{i}} \bar{n}_{\bar{i}} \bar\phi^{\bar{i}} K,_{\bar{i}} )\right].
\label{breakingterms1}
\eea
Taking a spinor derivative of the trace equations and adding implies the conservation equation for the supercurrent
\be
\partial_\mu V^\mu = iw \Gamma +i (DDS -\bar{D}\bar{D}\bar{S} ),
\ee
where the Ward-Takahashi identity functional differential operator $w \equiv D^\alpha w_\alpha -\bar{D}_{\dot\alpha} \bar{w}^{\dot\alpha}$.  

The identification of the component currents of the supercurrent can be obtained by expanding the superfields in terms of their component field structure.  At the same time the action also can be expanded in component fields.  Noether's theorem can be applied for each symmetry current and the current algebra can be used to define the SUSY tensorial central charge currents.  So doing the corresponding identification of the supercurrent components is secured.  This is done for the generalized Wess-Zumino model with canonical K\"ahler potential, using $\bar{i}=i$ when no ambiguity arises,
\be
K(\phi, \bar\phi) = \bar\phi^i \phi^i .
\ee
In this case the supercurrent is given by
\be
V_{\alpha\dot\alpha} = -\frac{8}{3}D_\alpha \phi^i \bar{D}_{\dot\alpha} \bar\phi^i + \frac{16}{3}i \phi^i \stackrel{\leftrightarrow}{\rlap{/}{\partial}}_{\alpha\dot\alpha} \bar\phi^i ,
\label{supercurrentsuperfield}
\ee
and the breaking terms in the trace identities are
\bea
S &=& -[ 2W +\sum_i n_i\phi^i W,_{i} +\bar{D}\bar{D} (\frac{2}{3}\bar\phi^i \phi^i +\sum_i n_i \bar\phi^i \phi^i)] \cr
\bar{S} &=& -[ 2\bar{W} +\sum_i n_i\bar\phi^i \bar{W},_{i} +{D}{D} (\frac{2}{3}\bar\phi^i \phi^i +\sum_i n_i \bar\phi^i \phi^i)] .
\label{breakingterms2}
\eea
which become for canonical scaling weight $n_i =-2/3$
\bea
S &=& -2[ W -\frac{1}{3}\phi^i W,_{i}] \cr
\bar{S} &=& -2 [ \bar{W} -\frac{1}{3}\bar\phi^i \bar{W},_{i} ].
\eea

The component Lagrangian and needed symmetry transformations of the fields are specified in Appendix B.  The $R$-symmetry current is 
\be
R^\mu =\frac{32}{3}i A^i \stackrel{\leftrightarrow}{\partial^\mu} \bar{A}^i -\frac{8}{3}\psi^i \sigma^\mu \bar\psi^i 
\ee
with conservation equation given as an algebraic identity
\be
\partial_\mu R^\mu = \delta^R {\cal L} -w^R(x) \Gamma ,
\label{Rconservation}
\ee
where the $R$ breaking terms arise from the variation of the Lagrangian
\bea
\delta^R {\cal L} &=& +4i \left[ F^i ( n W,_{ij} A^j + (n+2) W,_i ) -\frac{1}{4}\psi^i \psi^j ( nW,_{ijk} A^k + 2(n+1)W,_{ij} )\right]\cr
 & & -4i\left[ \bar{F}^i ( n \bar{W},_{ij} \bar{A}^j + (n+2) \bar{W},_i ) -\frac{1}{4}\bar\psi^i \bar\psi^j ( n\bar{W},_{ijk} \bar{A}^k + 2(n+1)\bar{W},_{ij} )\right] .\cr
 & & 
\eea
The scale and related $R$ weights of all fields have been taken to be the same, denoted by $R$ weight $n$.  The Ward-Takahashi identity functional differential operator for $R$ transformations is defined by $w^R(x) \equiv \sum_{\varphi}\delta^R \varphi^i (x) \delta/\delta \varphi^i (x)$, where $\varphi \in \{A, \psi, F, \bar{A}, \bar\psi, \bar{F}\}$ runs over all the component fields.  Technically the current conservation equations are to be viewed as composite operator insertions in one-particle irreducible Green's functions.  In the tree approximation considered here such an insertion reduces to equation (\ref{Rconservation}) with $\Gamma$ the tree approximation effective action (1-PI generating functional) which has the form of the classical action given in Appendix B equations (\ref{componentaction}) and (\ref{componentlag}).  That is any composite operator inserted 1-PI function reduces to just the operator in terms of sources, that is ${\cal O}(x)\Gamma = {\cal O}(x)$ in the Born approximation, such is the case with the composite Euler-Lagrange equations.  For the canonical case that $n=-2/3$ the $R$ variation operator is $w^R (x)= i[ \frac{2}{3}A(x)\delta/\delta A(x) -\frac{1}{3}\psi^\alpha (x) \delta/\delta \psi^\alpha (x) - \frac{2}{3}F(x) \delta/\delta F(x) + \cdots ]$ where the field index $i$ has been suppressed.  

In the SUSY case the Lagrangian is not invariant but transforms into a total derivative, hence, as in equation (\ref{Qcurrents1}), the Noether SUSY currents, defined here with the aid of the SUSY transformation Grassmann Weyl spinor parameters $\xi^\alpha$ and $\bar\xi^{\dot\alpha}$, are given by 
\bea
Q^\mu_{\rm N} (\xi) &=& 16 (\psi^i \sigma^\mu \bar\sigma^\rho \xi) \partial_\rho \bar{A}^i -4i \bar{W},_{i} \xi \sigma^\mu \bar\psi^i \cr
 &=&16\xi\psi^i \partial^\mu \bar{A}^i -16i \psi^i \sigma^{\mu\rho} \xi \partial_\rho \bar{A}^i -4i \bar{W},_i \xi \sigma^\mu \bar\psi^i \cr
\bar{Q}^\mu_{\rm N} (\bar\xi) &=& 16 (\bar\xi \bar\sigma^\rho \sigma^\mu  \bar\psi^i ) \partial_\rho {A}^i +4i {W},_{i} \psi^i \sigma^\mu \bar\xi  \cr
 &=&16\bar\xi \bar\psi^i \partial^\mu {A}^i -16i \bar\xi \bar\sigma^{\rho\mu} \bar\psi^i \partial_\rho {A}^i +4i {W},_i  \psi^i \sigma^\mu \bar\xi .
\eea
The SUSY currents are conserved
\bea
\partial_\mu Q^\mu_{\rm N} (\xi) &=& - w^Q (\xi) \Gamma \cr
\partial_\mu \bar{Q}^\mu_{\rm N} (\bar\xi) &=& - w^{\bar{Q}} (\bar\xi) \Gamma .
\eea
Using the Euler-Lagrange derivative with respect to the auxiliary fields in order to express the derivative of the superpotential in terms of them, modified SUSY currents are defined as (summation over the field index is suppressed)
\bea
Q^\mu_{\rm M} (\xi) &\equiv& Q^\mu_{\rm N} (\xi) -i \frac{\delta \Gamma}{\delta \bar{F}}(\xi \sigma^\mu \bar\psi) =16 (\psi \sigma^\mu \bar\sigma^\rho \xi) \partial_\rho \bar{A} -16i F(\xi \sigma^\mu \bar\psi) \cr
\bar{Q}^\mu_{\rm M} (\bar\xi) &\equiv& \bar{Q}^\mu_{\rm N} (\bar\xi) +i \frac{\delta \Gamma}{\delta {F}}(\psi \sigma^\mu \bar\xi) = 16 (\bar\xi \bar\sigma^\rho \sigma^\mu  \bar\psi ) \partial_\rho {A}^i +16i \bar{F}( \psi \sigma^\mu \bar\xi ) .
\eea
Now the conservation equations for the modified SUSY currents are those of the Noether SUSY currents with the addition of total derivatives of composite Euler-Lagrange equations
\bea
\partial_\mu Q^\mu_{\rm M} (\xi) &=& - w^Q (\xi) \Gamma -i\partial_\mu (\xi \sigma^\mu \bar\psi \frac{\delta \Gamma}{\delta \bar{F}})\cr
\partial_\mu \bar{Q}^\mu_{\rm M} (\bar\xi) &=& - w^{\bar{Q}} (\bar\xi) \Gamma +i\partial_\mu (\psi \sigma^\mu \bar\xi \frac{\delta \Gamma}{\delta F}).
\eea
Finally improved SUSY currents can be defined
\bea
Q^\mu (\xi) &\equiv& Q^\mu_M (\xi) +\partial_\rho Q^{\rho\mu}(\xi) \cr
\bar{Q}^\mu (\bar\xi) &\equiv& \bar{Q}^\mu_M (\bar\xi) +\partial_\rho \bar{Q}^{\rho\mu}(\bar\xi) ,
\eea
with the improvement terms
\bea
Q^{\rho\mu} (\xi) &=& \xi^\alpha Q_\alpha^{\rho\mu} = a (\psi \sigma^{\rho\mu}\xi ) \bar{A} \cr
\bar{Q}^{\rho\mu} (\xi) &=& \bar\xi_{\dot\alpha} \bar{Q}^{\rho\mu\dot\alpha} = \bar{a} (\bar\xi \bar\sigma^{\rho\mu} \bar\psi) {A} .
\eea
The improvements are chosen to make the trace \lq\lq soft" for a renormalizable superpotential and the superconformal currents expressed as the simple space-time moment of the SUSY currents, as seen in equations (\ref{impS}) and (\ref{superconformalmoment}); hence $a=-\frac{32}{3}i$ and $\bar{a}= +\frac{32}{3}i$.  Thus the improved SUSY currents have the component field form
\bea
Q^\mu (\xi) &=& 16 (\xi\psi) \partial^\mu \bar{A} -16i F(\xi \sigma^\mu \bar\psi) +\frac{32}{3}i (\partial_\rho \psi \sigma^{\mu\rho} \xi ) \bar{A} -\frac{16}{3}i (\psi \sigma^{\mu\rho} \xi ) \partial_\rho \bar{A} \cr
\bar{Q}^\mu (\bar\xi) &=& 16 (\bar\xi\bar\psi) \partial^\mu {A} +16i \bar{F}( \psi\sigma^\mu \bar\xi) +\frac{32}{3}i ( \bar\xi\bar\sigma^{\rho\mu}  \partial_\rho \bar\psi) {A} -\frac{16}{3}i (\bar\xi\bar\sigma^{\rho\mu} \bar\psi  ) \partial_\rho {A} ,\cr
 & & 
\label{modimpQ}
\eea
and conservation equations
\bea
\partial_\mu Q^\mu (\xi) &=& -w^Q (\xi) \Gamma -i \partial_\mu ( \xi \sigma^\mu \bar\psi \frac{\delta \Gamma}{\delta \bar{F}} )\cr
\partial_\mu \bar{Q}^\mu (\bar\xi) &=& -w^{\bar{Q}} (\bar\xi) \Gamma +i \partial_\mu ( \psi \sigma^\mu \bar\xi \frac{\delta \Gamma}{\delta {F}} )
\eea
and \lq\lq soft" trace equations
\bea
(Q^\mu \sigma_\mu)_{\dot\alpha} &=& 3i[a +\frac{32}{3}i ] (\psi \sigma^\mu )_{\dot\alpha} \partial_\mu \bar{A}  +16i [ \bar{W},_i - \frac{3}{64}ia \bar{A}^j \bar{W},_{ij} ] \bar\psi^i_{\dot\alpha} + \frac{3}{8} a \bar{A}^i \frac{\delta \Gamma}{\delta \bar\psi^{i\dot\alpha}} \cr
 &=&+16i [ \bar{W},_i - \frac{1}{2} \bar{A}^j \bar{W},_{ij} ] \bar\psi^i_{\dot\alpha} -4i \bar{A}^i \frac{\delta \Gamma}{\delta \bar\psi^{i\dot\alpha}}\cr
(\sigma_\mu \bar{Q}^\mu )_{\alpha} &=& -3i[\bar{a} -\frac{32}{3}i ] (\sigma^\mu \bar\psi )_{\alpha} \partial_\mu {A} -16i [ {W},_i +\frac{3}{64} i\bar{a} {W},_{ij}{A}^j ] \psi^i_{\alpha} - \frac{3}{8} \bar{a} \frac{\delta \Gamma}{\delta \psi^{i\alpha}}{A}^i \cr
 &=&-16i [ {W},_i - \frac{1}{2} {W},_{ij}{A}^j ] \psi^i_{\alpha} - 4i \frac{\delta \Gamma}{\delta \psi^{i\alpha}}{A}^i .
\eea

The Noether energy-momentum tensor can be found using the space-time translation intrinsic variation of each component field, $\delta^P_\mu \varphi = \partial_\mu \varphi$, and the fact that the variation of the Lagrangian is its derivative, $\delta^P_\mu {\cal L} =  \partial_\mu {\cal L}$.  These yield the Noether energy-momentum tensor
\be
T^{\mu\nu}_{\rm N} = 16[ \partial^\mu A \partial^\nu \bar{A} + \partial^\nu A \partial^\mu \bar{A}] -4i [\partial^\nu \psi \sigma^\mu \bar\psi - \psi \sigma^\mu \partial^\nu \bar\psi ] - g^{\mu\nu} {\cal L} 
\ee
with conservation equation
\be
\partial_\mu T^{\mu\nu}_{\rm N} = -w^{P\nu} (x) \Gamma 
\ee
where the translation Ward-Takahashi identity operator is given by
\be
w^P_\mu (x) = \partial_\mu A \frac{\delta}{\delta A} + \partial_\mu \psi^\alpha \frac{\delta}{\delta \psi^\alpha} + \cdots .
\ee
A symmetric energy-momentum tensor can be defined by means of the Belinfante improvement procedure
\be
T^{\mu\nu}_{\rm B} = T^{\mu\nu}_{\rm N} + \partial_\rho G^{\rho\mu\nu}_{\rm B} 
\ee
with the improvement term chosen to symmetrize the energy-momentum tensor, modulo Euler-Lagrange terms,
\be
G^{\rho\mu\nu}_{\rm B} = -16 \epsilon^{\mu\nu\rho\sigma} (\psi \sigma_\sigma \bar\psi ) .
\ee
Exploiting the fermion field equations as algebraic identities, the Belinfante energy-momentum tensor is obtained
\bea
T^{\mu\nu}_{\rm B} &=& 16[ \partial^\mu A \partial^\nu \bar{A} + \partial^\nu A \partial^\mu \bar{A}] +2i [\psi \sigma^\nu \stackrel{\leftrightarrow}{\partial^\mu} \bar\psi + \psi \sigma^\mu \stackrel{\leftrightarrow}{\partial^\nu} \bar\psi ] - g^{\mu\nu} {\cal L} \cr
 & & \qquad\qquad -2i [\psi \sigma^{\mu\nu} \frac{\delta\Gamma}{\delta \psi} +\frac{\delta \Gamma}{\delta \bar\psi} \bar\sigma^{\mu\nu} \bar\psi ] .
\eea
The conservation equation remains the same $\partial_\mu T^{\mu\nu}_{\rm B} = \partial_\mu T^{\mu\nu}_{\rm N} = -w^{P\nu} (x) \Gamma$.  The trace of the Belinfante tensor however is not \lq\lq soft".  A further improved tensor with a soft trace can be obtained by adding yet another trivially conserved term to $T^{\mu\nu}_{\rm B}$
\be
T^{\mu\nu}_{\rm I} = T^{\mu\nu}_{\rm B} +\xi (g^{\mu\nu} \partial^2 -\partial^\mu \partial^\nu ) (32 A\bar{A}) ,
\ee
where the \lq\lq soft" trace condition yields $\xi = 1/6$.  

The final form of the conserved, symmetric and \lq\lq soft" trace energy-momentum tensor is obtained by once more modifying the current by subtracting composite Euler-Lagrange equations from the improved tensor
\bea
T^{\mu\nu} &=& T^{\mu\nu}_{\rm I} - E^{\mu\nu} \cr
 &=& 16[ \partial^\mu A \partial^\nu \bar{A} + \partial^\nu A \partial^\mu \bar{A}] - g^{\mu\nu} 16 \partial_\rho A \partial^\rho \bar{A} +\frac{1}{3}(g^{\mu\nu} \partial^2 - \partial^\mu \partial^\nu ) 16(A\bar{A} )  \cr
 & & \qquad\qquad + g^{\mu\nu} 16 F \bar{F} +2i [\psi \sigma^\nu \stackrel{\leftrightarrow}{\partial}^\mu \bar\psi + \psi \sigma^\mu \stackrel{\leftrightarrow}{\partial}^\nu \bar\psi ] ,
\eea
where the composite Euler-Lagrange terms are given by
\be
E^{\mu\nu} = -g^{\mu\nu} \left[ F\frac{\delta \Gamma}{\delta F} + \bar{F}\frac{\delta \Gamma}{\delta \bar{F}} + \frac{1}{2}\left( \psi \frac{\delta \Gamma}{\delta \psi} -\frac{\delta \Gamma}{\delta \bar{\psi}} \bar\psi \right)\right] -2i \left[ \psi \sigma^{\mu\nu} \frac{\delta \Gamma}{\delta \psi} + \frac{\delta \Gamma}{\delta \bar\psi} \bar\sigma^{\mu\nu} \bar\psi \right] .
\ee
The energy-momentum tensor is conserved 
\be
\partial_\mu T^{\mu\nu} = \partial_\mu T^{\nu\mu} = \partial_\mu T^{\mu\nu}_{\rm N} -\partial_\mu E^{\mu\nu} = -w^{P\nu} (x) - \partial_\mu E^{\mu\nu} 
\ee
and has the \lq\lq soft" trace
\bea
T^\lambda_{~\lambda} &=& 16( A\partial^2 \bar{A} + \partial^2 A \bar{A} ) + 64 F\bar{F} + 4i \psi \stackrel{\leftrightarrow}{\rlap{/}{\partial}} \bar\psi \cr
 &=& T^\lambda_{{\rm I}~\lambda} + 4 \left[ F\frac{\delta \Gamma}{\delta F} + \bar{F}\frac{\delta \Gamma}{\delta \bar{F}} + \frac{1}{2}\left( \psi \frac{\delta \Gamma}{\delta \psi} -\bar\psi \frac{\delta \Gamma}{\delta \bar\psi} \right) \right] \cr
 &=& 8 [W,_i -\frac{1}{2}W,_{ij} A^j ] F^i +8 [ \bar{W},_i -\frac{1}{2}\bar{W},_{ij} \bar{A}^j ] \bar{F}^i \cr
 & & - [ W,_{ij} -W,_{ijk} A^k ] \psi^i \psi^j - [ \bar{W},_{ij} -\bar{W},_{ijk} \bar{A}^k ] \bar\psi^i \bar\psi^j \cr
 & & \left[ -\left( A\frac{\delta\Gamma}{\delta A}+\bar{A}\frac{\delta\Gamma}{\delta \bar{A}}\right) +\frac{1}{2} \left( \psi \frac{\delta\Gamma}{\delta \psi} -\bar\psi \frac{\delta\Gamma}{\delta\bar\psi}\right) + 2\left( F \frac{\delta\Gamma}{\delta F} +\bar{F} \frac{\delta\Gamma}{\delta \bar{F}}\right)\right] .
\eea

Finally, the topological charge currents, denoted $\zeta^\mu_{\alpha\beta}$ and $\bar\zeta^\mu_{\dot\alpha\dot\beta}$, and the SUSY tensorial central charge currents, denoted $Z^\mu_{\alpha\beta}$ and $\bar{Z}^\mu_{\dot\alpha\dot\beta}$, are found by considering the SUSY variations of the modified and improved SUSY currents in equation (\ref{modimpQ})
\bea
\delta^Q_\alpha Q^\mu_\beta + \delta^Q_\beta Q^\mu_\alpha &=& Z^\mu_{\alpha\beta}= \zeta^\mu_{\alpha\beta} -8i\sigma^{\mu\nu}_{\alpha\beta}\left[ \frac{\delta\Gamma}{\delta\bar{F}} \partial_\nu \bar{A} -\frac{1}{3}\partial_\nu \left(\frac{\delta\Gamma}{\delta\bar{F}} \bar{A}\right)\right] \cr
\delta^{\bar{Q}}_{\dot\alpha} \bar{Q}^\mu_{\dot\beta} + \delta^{\bar{Q}}_{\dot\beta} \bar{Q}^\mu_{\dot\alpha} &=& \bar{Z}^\mu_{\dot\alpha\dot\beta}= \bar\zeta^\mu_{\dot\alpha\dot\beta} -8i\bar\sigma^{\mu\nu}_{\dot\alpha\dot\beta}\left[ \frac{\delta\Gamma}{\delta {F}} \partial_\nu {A} -\frac{1}{3}\partial_\nu \left(\frac{\delta\Gamma}{\delta{F}} {A}\right)\right] ,
\eea
where the topological currents are given by
\bea
\zeta^\mu_{\alpha\beta} &=& -32i \sigma^{\mu\nu}_{\alpha\beta} \partial_\nu \left(\bar{W} -\frac{1}{3}\bar{W},_i \bar{A}^i \right) \cr
\bar\zeta^\mu_{\dot\alpha\dot\beta} &=& -32i \bar\sigma^{\mu\nu}_{\dot\alpha\dot\beta} \partial_\nu \left({W} -\frac{1}{3}{W},_i {A}^i \right) .
\eea
As in the case of the Noether version of these currents, on-shell they are equal, inserted in Green's functions they differ by composite Euler-Lagrange equations.  The SUSY tensorial central charge currents obey the conservation equations with Ward-Identity functional differential operator divergences
\bea
\partial_\mu Z^\mu_{\alpha\beta} &=& -8i \sigma^{\mu\nu}_{\alpha\beta} (\partial_\mu \frac{\delta\Gamma}{\delta \bar{F}})(\partial_\nu \bar{A}) \cr
\partial_\mu \bar{Z}^\mu_{\dot\alpha\dot\beta} &=& -8i \bar\sigma^{\mu\nu}_{\dot\alpha\dot\beta} (\partial_\mu \frac{\delta\Gamma}{\delta F})(\partial_\nu A).
\eea
This can be interpreted as the invariance of the theory under a local tensorial symmetry transformation involving only the auxiliary fields as found in references \cite{Fujikawa:2003gi}, \cite{Shizuya:2003vm}.  The conserved SUSY tensorial central charge currents correspond to a local symmetry transformation of the auxiliary fields
\bea
\delta^Z (\zeta, \bar\zeta) F &=& -4i \bar\sigma^{\mu\nu}_{\dot\alpha\dot\beta} (\partial_\mu \bar\zeta^{\dot\alpha\dot\beta})(\partial_\nu A) \cr
\delta^Z (\zeta, \bar\zeta) \bar{F} &=& -4i \sigma^{\mu\nu}_{\alpha\beta} (\partial_\mu \zeta^{\alpha\beta})(\partial_\nu \bar{A}) ,
\eea
all other fields being invariant and the local central charge transformation parameters are $\zeta_{\alpha\beta} (x)$ and $\bar\zeta_{\dot\alpha\dot\beta} (x)$.  

Expanding the superfield expression for the supercurrent, equation (\ref{supercurrentsuperfield}), in terms of the component fields, the identification of its components in terms of the improved currents can be made
\bea
V^\mu &=& C^\mu + i\theta^\alpha \chi^\mu_\alpha -i \bar\theta_{\dot\alpha} \bar\chi^{\dot\alpha\mu} + \frac{1}{2} \theta^2 M^\mu +\frac{1}{2}\bar\theta^2 \bar{M}^\mu + 2 \theta \sigma_\nu \bar\theta v^{\mu\nu} \cr
 & & \qquad\qquad + \frac{1}{2}\theta^2 \bar\theta_{\dot\alpha} \bar\lambda^{\dot\alpha\mu} +\frac{1}{2}\bar\theta^2 \theta^\alpha \lambda^\mu_\alpha +\frac{1}{4}\theta^2 \bar\theta^2 D^\mu ,
\eea
where the components are given by
\bea
C^\mu&=& R^\mu \cr
 & & \cr
\chi^\mu_{\alpha} &=& \left[ Q^\mu -\frac{1}{3}Q^\nu \sigma_\nu \bar\sigma^\mu \right]_\alpha \cr
 & & \cr
\bar\chi^\mu_{\dot\alpha} &=& \left[ \bar{Q}^\mu -\frac{1}{3}Q^\nu \bar\sigma^\mu \sigma_\nu \right]_{\dot\alpha} \cr
 & & \cr
M^\mu &=& \frac{1}{12} \sigma^{\mu\nu\beta}_\alpha Z_{\nu\beta}^{~~~\alpha} \cr
 & & \cr
\bar{M}^\mu &=& -\frac{1}{12} \bar\sigma^{\mu\nu\dot\alpha}_{~~~\dot\beta} \bar{Z}^{\dot\beta}_{\nu~~\dot\alpha} \cr
 & & \cr
v^{\mu\nu} &=& \left[ T^{\mu\nu} -\frac{1}{3}g^{\mu\nu} T^\lambda_{~\lambda}\right] +\frac{1}{4}\epsilon^{\mu\nu\rho\sigma} \partial_\rho R_\sigma +\frac{i}{4}\left[ \psi\sigma^{\mu\nu} \frac{\delta\Gamma}{\delta \psi} +\frac{\delta\Gamma}{\delta \bar\psi} \bar\sigma^{\mu\nu} \bar\psi \right] \cr
 & & \cr
\bar\lambda^\mu &=& \left[ \partial^\mu \bar\sigma^\nu -i\epsilon^{\mu\nu\rho\sigma} \bar\sigma_\rho \partial_\sigma \right] \left[ Q^\mu (x)- \frac{1}{3}Q^\rho (x)\sigma_\rho \bar\sigma^\mu \right] \cr
 & & \qquad +(\bar\sigma^\mu \psi ) \frac{\delta\Gamma}{\delta A} +6i \frac{\delta\Gamma}{\delta\bar\psi} \partial^\mu \bar{A} + 2 \frac{\delta\Gamma}{\delta \bar\psi} \bar\sigma^{\mu\nu}\partial_\nu \bar{A} \cr
 & & \qquad\qquad+ 2 F\bar\sigma^\mu \frac{\delta\Gamma}{\delta \psi} -3i  (\partial^\mu \frac{\delta\Gamma}{\delta \bar{F}})\bar\psi + (\partial_\nu \frac{\delta\Gamma}{\delta\bar{F}})(\bar\sigma^{\mu\nu} \bar\psi) \cr
 & & \cr
\lambda^\mu &=& \left[ \partial^\mu \sigma^\nu -i\epsilon^{\mu\nu\rho\sigma} \sigma_\rho \partial_\sigma \right]\left[ \bar{Q}^\mu (x)-\frac{1}{3}\bar\sigma^\mu \sigma^\rho \bar{Q}_\rho (x)\right]\cr
 & & \qquad -\frac{\delta\Gamma}{\delta \bar{A}}(\sigma^\mu \bar\psi )  +6i \partial^\mu {A}\frac{\delta\Gamma}{\delta \psi} - 2 \partial_\nu {A} \sigma^{\mu\nu}\frac{\delta\Gamma}{\delta \psi} \cr
 & & \qquad\qquad - 2 ( \frac{\delta\Gamma}{\delta \bar\psi \bar\sigma^\mu})\bar{F} +3i \psi (\partial^\mu \frac{\delta\Gamma}{\delta {F}}) - (\sigma^{\mu\nu} \psi)(\partial_\nu \frac{\delta\Gamma}{\delta{F}}) \cr
 & & \cr
D^\mu &=& \left[ \partial^2 g^{\mu\nu} -2 \partial^\mu \partial^\nu \right] R_\nu (x) \cr
 & & \cr
 & &\qquad +2i \psi \stackrel{\leftrightarrow}{\partial^\mu} \frac{\delta\Gamma}{\delta \psi} -2i \frac{\delta\Gamma}{\delta\bar\psi}\stackrel{\leftrightarrow}{\partial^\mu} \bar\psi \cr
 & & \cr
 & &\qquad\qquad +4i\left[ -\partial^\mu A \frac{\delta\Gamma}{\delta A} + F \partial^\mu \frac{\delta\Gamma}{\delta F} +\partial^\mu \bar{A} \frac{\delta\Gamma}{\delta \bar{A}} -\bar{F} \partial^\mu \frac{\delta\Gamma}{\delta \bar{F}}\right] .
\eea
As usual, on-shell the Euler-Lagrange terms vanish and the supercurrent components are given in terms of the super-Poincar\'e currents as in equation (\ref{supercurrentcomp}).
As in equations (\ref{currents}) and (\ref{tencentcurrents}) and (\ref{Scurrents}), the extended tensorial central charge superconformal currents can all be obtained from the supercurrent with their (non-)conservation equations following from the generalized traces of the supercurrent, $D^\alpha V_{\alpha\dot\alpha}$ and $\bar{D}^{\dot\alpha} V_{\alpha\dot\alpha}$.

\newsection{Radiative Corrections}

A rigorous treatment of the radiative corrections to the quantization of the generalized Wess-Zumino model about a domain wall solution is beyond the scope of this paper \cite{Shifman:1998zy}, \cite{Shizuya:2005th}-\cite{Fujikawa:2003gi}.  However, based on some general assumptions, the conjectured results of a renormalization program will be discussed.  The trace of the supercurrent as an algebraic identity is expressed in equation (\ref{supercurrenttrace}) where now the distinction between the Lagrangian and its action, denoted $\Gamma_0$, and the effective action which is the generator of 1-PI functions, denoted $\Gamma$, must be made so that the trace equations become the algebraic identities
\bea
-2(w_\alpha \Gamma_0)_{\rm hard} \Gamma &=& \bar{D}^{\dot\alpha} V_{\alpha\dot\alpha}\Gamma +2D_\alpha S_{\rm hard} \Gamma \cr
-2(\bar{w}_{\dot\alpha} \Gamma_0)_{\rm hard} \Gamma &=& {D}^{\alpha} V_{\alpha\dot\alpha}\Gamma +2\bar{D}_{\dot\alpha} \bar{S}_{\rm hard} \Gamma,
\label{hardeom}
\eea
where the renormalized (\lq\lq hard") composite operators, including the supercurrent, are understood to be inserted in the 1-PI function generating functional $\Gamma$.  The quantum action principle applied to the composite Euler-Lagrange equations on the left hand side of these equations as well as the renormalization of the operators in equation (\ref{hardeom}) will determine the radiative corrections to this identity.  Assuming a completely invariant renormalization scheme exists such as a BPHZ momentum superspace subtraction procedure \cite{Clark:1977pq} with perhaps an auxiliary subtraction parameter and/or a modified subtraction algorithm to account for the domain wall background quantization, the composite Euler-Lagrange equations would maintain their naive form.  This results in the inserted Euler-Lagrange equations being replaced with just the Ward-Takahashi functional differential operator acting on the effective action, hence the renormalized trace identities in this scheme take the simple form
\bea
-2w_\alpha  \Gamma &=& \bar{D}^{\dot\alpha} V_{\alpha\dot\alpha}\Gamma +2D_\alpha S_{\rm hard} \Gamma \cr
-2\bar{w}_{\dot\alpha} \Gamma &=& {D}^{\alpha} V_{\alpha\dot\alpha}\Gamma +2\bar{D}_{\dot\alpha} \bar{S}_{\rm hard} \Gamma~~,
\eea
where the supercurrent $V_{\alpha\dot\alpha}$ and the breaking terms $S_{\rm hard}$ and $\bar{S}_{\rm hard}$ are understood to be \lq\lq BPHZ" renormalized operators.  The breaking terms $S_{\rm hard}$ and $\bar{S}_{\rm hard}$ have their naive form as in equations (\ref{breakingterms1}) and (\ref{breakingterms2}) however there is an over subtraction of the super renormalizable terms appearing in $S_{\rm hard}$ and $\bar{S}_{\rm hard}$; this is the price paid in order to maintain the naive form of the equations of motion.  The over subtracted, and therefore hard, terms are related to the corresponding minimally renormalized, and hence soft, terms by the addition of all power counting hard composite operators that can be constructed maintaining the symmetries respected by this idealized BPHZ scheme.  For example, the over subtracted chiral mass term in the original Wess-Zumino model, $N_{\rm hard}[\phi^2]$, is related to the minimally subtracted mass term, $N_{\rm soft}[\phi^2]$, by the addition of all dimension 3 chiral composite operators through use of the Zimmermann identity \cite{Zimmermann}, \cite{Clark:1977pq}, \cite{Clark:1978jx}
\be
N_{\rm hard}[\phi^2] = N_{\rm soft}[\phi^2] + t N_{\rm hard}[\phi^3] +r N_{\rm hard}[\bar{D}\bar{D} \phi \bar\phi] + v N_{\rm hard}[\partial^2 \phi] +w N_{\rm hard}[\bar{D}\bar{D} \bar\phi^2] ,
\ee
where the $r$, $t$, $v$ and $w$ are $O(\hbar)$ coefficients.  The first coefficient can be shown to vanish due to the no renormalization theorem for the superpotential and hence a conserved $R$ symmetry, $t=0$ \cite{Clark:1978jx}.  The remaining coefficients can be consistently absorbed into a renormalization of the now modified supercurrent $V^\prime_{\alpha\dot\alpha}$ so that the trace equations become
\bea
-2w_\alpha  \Gamma &=& \bar{D}^{\dot\alpha} V_{\alpha\dot\alpha}^\prime\Gamma +2D_\alpha S^\prime ~\Gamma \cr
-2\bar{w}_{\dot\alpha} \Gamma &=& {D}^{\alpha} V^\prime_{\alpha\dot\alpha}\Gamma +2\bar{D}_{\dot\alpha} \bar{S}^\prime ~\Gamma~~,
\eea
with the breaking terms having the form
\bea
S^\prime &=& -\left[ 2W +\sum_i n_i\phi^i W,_{i} +\bar{D}\bar{D} \sum_i  \left(\frac{2}{3} + n_i+r_i \right)\bar\phi^i \phi^i \right]_{\rm minimum} \cr
\bar{S}^\prime &=& -\left[ 2\bar{W} +\sum_i n_i \bar\phi^i \bar{W},_{i} +{D}{D} \sum_i  \left(\frac{2}{3} + n_i+r_i \right)\bar\phi^i \phi^i\right]_{\rm minimum} ,
\label{Sprimekinetic}
\eea
where the \lq\lq minimum" subscript indicates that each term in the brackets is now minimally subtracted.  

The renormalized bilinear field equations can be used to express the trace equation in terms of the anomalous dimension of the superfields.  Inserting the bilinear equations of motion as an algebraic identity,
\bea
\phi^i \frac{\delta \Gamma_0}{\delta \phi^i} &=&  \bar{D}\bar{D} \phi^i \bar\phi^i + \phi^i W,_i \cr
\bar\phi^i \frac{\delta \Gamma_0}{\delta \bar\phi^i} &=& {D}{D} \phi^i \bar\phi^i + \bar\phi^i \bar{W,}_i ,
\eea
into the effective action $\Gamma$ and applying the \lq\lq BPHZ" renormalized quantum action principle yields the renormalized counting density identities
\bea
\phi^i \frac{\delta}{\delta \phi^i}\Gamma &=&  \left[ \bar{D}\bar{D} \phi^i \bar\phi^i + \phi^i W,_i \right]_{\rm hard} \Gamma\cr
\bar\phi^i \frac{\delta}{\delta \bar\phi^i}\Gamma &=& \left[{D}{D} \phi^i \bar\phi^i + \bar\phi^i \bar{W,}_i \right]_{\rm hard} \Gamma  .
\eea
Exploiting the Zimmermann identity to relate the over subtracted to the minimally subtracted terms here, as in the trace equations into which they are substituted, results in a redefinition of the supercurrent coefficients and the appearance of the renormalization group anomalous dimension of the fields (it is related to the $r$ coefficient) in the scale and $R$ weights of the Ward-Takahashi functional differential operator and the appearance of an anomalous dimension correction to the breaking terms.  In short the familiar results for the renormalization of the trace of the supercurrent are assumed to hold in the spontaneous breaking domain wall case as well as in its absence \cite{Clark:1978jx}, \cite{Konishi:1983hf}.  With a further renormalization of the supercurrent, denoted $V_{\alpha\dot\alpha}^{\rm ren.}$, similar in form to $V_{\alpha\dot\alpha}^\prime$, the renormalized trace equations become
\bea
-2w_\alpha^{(\gamma)}  \Gamma &=& \bar{D}^{\dot\alpha} V_{\alpha\dot\alpha}^{\rm ren.}~\Gamma +2D_\alpha S^{\rm ren.} ~\Gamma \cr
-2\bar{w}_{\dot\alpha}^{(\gamma)} \Gamma &=& {D}^{\alpha} V_{\alpha\dot\alpha}^{\rm ren.}~\Gamma +2\bar{D}_{\dot\alpha} \bar{S}^{\rm ren.} ~\Gamma~~,
\eea
where the anomalous dimension Ward-Takahashi identity operators are given by
\bea
w_\alpha^{(\gamma)} \Gamma &=& 2(D_\alpha \phi^i ) \frac{\delta \Gamma}{\delta \phi^i} -\frac{2}{3}\sum_{{i}}(1+\gamma_i) D_\alpha\left(\phi^i \frac{\delta \Gamma}{\delta \phi^i} \right)\cr
\bar{w}_{\dot\alpha}^{(\gamma)} \Gamma &=& 2(\bar{D}_{\dot\alpha} \bar\phi^{{i}} ) \frac{\delta \Gamma}{\delta \bar\phi^{{i}}} + \frac{2}{3}\sum_{{i}} (1+\gamma_i) \bar{D}_{\dot\alpha}\left(\bar\phi^{{i}} \frac{\delta \Gamma}{\delta \bar\phi^{{i}}} \right).
\eea
With appropriate normalization the symmetry breaking terms become
\bea
S^{\rm ren.} &=& -\left[ 2W +\sum_i n_i(1+\gamma_i)\phi^i W,_{i} +\bar{D}\bar{D} \sum_i\left(\frac{2}{3}+ n_i   \right)\bar\phi^i \phi^i\right]_{\rm minimum} \cr
\bar{S}^{\rm ren.} &=& -\left[ 2\bar{W} +\sum_i n_i(1+\gamma_i)\bar\phi^i \bar{W},_{i} +{D}{D} \sum_i \left(\frac{2}{3} + n_i   \right)\bar\phi^i \phi^i\right]_{\rm minimum} ,
\eea
which, for canonical scaling weight $n_i =-2/3$, become 
\bea
S^{\rm ren.} &=& -2[ W -\frac{1}{3}\sum_i (1+\gamma_i) \phi^i W,_{i} ]_{\rm minimum} \cr
\bar{S}^{\rm ren.} &=& -2 [ \bar{W} -\frac{1}{3}\sum_i (1+\gamma_i)\bar\phi^i \bar{W},_{i} ]_{\rm minimum} .
\eea
The radiative corrections to the kinetic terms in equation (\ref{Sprimekinetic}) have been expressed as anomalous dimension corrections to the improvement terms.  Thus for the case of a domain wall solution to the BPS equations, $\frac{dA(z)}{dz} = W^\prime (A(z))$, and hence the field equations (\ref{fieldequations}), the contribution of the breaking terms involving the anomalous dimension to the topological central charge vanish either in the form of $S^\prime$, $\bar{S}^\prime$ as $\bar{D}\bar{D} (\bar\phi \phi)$ and $DD(\bar\phi \phi)$ terms or in the form $S^{\rm ren.}$, $\bar{S}^{\rm ren.}$ as $\phi^i W,_i$ and $\bar\phi \bar{W},_i$ terms, as will be seen in the following.

Returning to the expression for the SUSY tensorial central charge current, equation (\ref{currents}), they are given by the generalized trace of the supercurrent
\bea
\hat{Z}^\mu_{\alpha\beta} &=& +2i\left[ \bar\sigma^{\mu\dot\gamma}_\alpha D_\beta +\bar\sigma^{\mu\dot\gamma}_\beta D_\alpha \right] D^\gamma 
V_{\gamma\dot\gamma}^{\rm ren.} \cr
 &=& +16i\sigma^{\mu\nu}_{\alpha\beta} \partial_\nu \bar{S}^{\rm ren.} -4i [ \bar\sigma^{\mu\dot\gamma}_{~~~\alpha} D_\beta +\bar\sigma^{\mu\dot\gamma}_{~~~\beta} D_\alpha ] \bar{w}_{\dot\gamma}^{(\gamma)} \Gamma \cr
 &=& \hat\zeta^\mu_{\alpha\beta}-4i [ \bar\sigma^{\mu\dot\gamma}_{~~~\alpha} D_\beta +\bar\sigma^{\mu\dot\gamma}_{~~~\beta} D_\alpha ] \bar{w}_{\dot\gamma}^{(\gamma)} \Gamma \cr
 & & \cr
\hat{\bar{Z}}^\mu_{\dot\alpha\dot\beta} &=& -2i\left[ \bar{D}_{\dot\alpha} \bar\sigma^{\mu\dot\gamma}_{\dot\beta}  +\bar{D}_{\dot\beta} \bar\sigma^{\mu\dot\gamma}_{\dot\alpha}\right] \bar{D}^{\dot\gamma} 
V_{\gamma\dot\gamma}^{\rm ren.} \cr
 &=& +16i\bar\sigma^{\mu\nu}_{\dot\alpha\dot\beta} \partial_\nu {S}^{\rm ren.} -4i [ \bar{D}_{\dot\alpha}\bar\sigma^{\mu\gamma}_{\dot\beta}  +\bar{D}_{\dot\beta}\bar\sigma^{\mu\gamma}_{\dot\alpha}] {w}_{\gamma}^{(\gamma)}\Gamma \cr
 &=& \hat{\bar{\zeta}}^\mu_{\dot\alpha\dot\beta}-4i [ \bar{D}_{\dot\alpha}\bar\sigma^{\mu\gamma}_{\dot\beta}  +\bar{D}_{\dot\beta}\bar\sigma^{\mu\gamma}_{\dot\alpha}] {w}_{\gamma}^{(\gamma)}\Gamma .
\eea
Thus the topological currents are identified as
\bea
\hat\zeta^\mu_{\alpha\beta} &=& +16i\sigma^{\mu\nu}_{\alpha\beta} \partial_\nu \bar{S}^{\rm ren.} \cr
\hat{\bar{\zeta}}^\mu_{\dot\alpha\dot\beta} &=& +16i\bar\sigma^{\mu\nu}_{\dot\alpha\dot\beta} \partial_\nu {S}^{\rm ren.} .
\eea
Again it is noted that for a domain wall solution to the BPS equations and hence the field equations (\ref{fieldequations}), the anomalous dimension contributions as well as the improved current contributions to the topological charge  vanish and the value of the topological charge remains unrenormalized \cite{Chibisov:1997rc}.  Explicitly, the topological currents are
\bea
\hat\zeta^\mu_{\alpha\beta} &=& -32i\sigma^{\mu\nu}_{\alpha\beta} \partial_\nu [ \bar{W} -\frac{1}{3}\sum_i (1+\gamma_i)\bar\phi^i \bar{W},_{i}] \cr
\hat{\bar{\zeta}}^\mu_{\dot\alpha\dot\beta} &=& -32i\bar\sigma^{\mu\nu}_{\dot\alpha\dot\beta} \partial_\nu [ W -\frac{1}{3}\sum_i (1+\gamma_i)\phi^i W,_{i}] .
\eea
In the simplest single field Wess-Zumino model with superpotential $W= m^2 \phi -\frac{\lambda}{3}\phi^3$, the domain wall solution in the $x$-$y$ plane $\phi = \frac{m}{\sqrt{\lambda}} \tanh{m\sqrt{\lambda}z} + \theta^2 \frac{m^2}{4\cosh^2{m\sqrt{\lambda}z}}$ results in the topological currents
\bea
\hat\zeta^\mu_{\alpha\beta} &=& -32i \sigma^{\mu 3}_{\alpha\beta} \partial_z \left(\bar{W} -\frac{1}{3}(1+\gamma) \bar\phi \bar{W}^\prime\right)=64im^2 \sigma^{\mu 3}_{\alpha\beta} \partial_z \left(\bar\phi\left[ \frac{2}{3}-\frac{1}{3}\gamma(1- \frac{\lambda}{m^2}\bar\phi^2)\right]\right) \cr 
\hat{\bar{\zeta}}^\mu_{\dot\alpha\dot\beta} &=& -32i \bar\sigma^{\mu 3}_{\dot\alpha\dot\beta} \partial_z \left(W-\frac{1}{3}(1+\gamma) \bar\phi \bar{W}^\prime\right)=  64im^2 \bar\sigma^{\mu 3}_{\dot\alpha\dot\beta} \partial_z \left(\phi\left[ \frac{2}{3}-\frac{1}{3}\gamma(1- \frac{\lambda}{m^2}\phi^2)\right]\right). \cr 
 & & 
\eea
Hence the topological charge density, which is equal to the BPS domain wall tension, is secured, as indicated by equation (\ref{topocharge}),
\be
\zeta \equiv \frac{1}{16}\int_{-\infty}^{+\infty} dz \hat\zeta^0_{11} = \frac{8}{3}\frac{m^3}{\sqrt{\lambda}}~~,
\ee
likewise using the $\hat{\bar{\zeta}}^\mu_{\dot\alpha\dot\beta}$ current.  

Finally the values of the conformal tensorial central charges evaluated on the domain wall are given in terms of the breaking terms $S^{\rm ren.}$ and $\bar{S}^{\rm ren.}$.  The conformal tensorial central charge currents are not trivially conserved and the conformal tensorial central charges receive radiative corrections.  As seen from equation (\ref{tencentcurrents}) the currents depend on the anomalous dimensions of the superfields where $d=(1+\gamma)$ for each superfield.  Also their non-conservation equations are given by equation (\ref{centralcurrentsdivergence}) with $d=(1+\gamma)$.  The $Y_{\alpha\dot\alpha}$ and $\bar{Y}_{\alpha\dot\alpha}$ conformal tensorial central charges that are constructed from the first space-time moment of the SUSY tensorial central charge currents, equation (\ref{tencentcurrents}), in fact receive no radiative corrections since as in the SUSY central charge case the anomalous dimension terms vanish upon integration across the domain wall.  From equation (\ref{tencentcurrents}) the conformal central charge current is given by
\be
Y^\mu_{\alpha\dot\alpha} = 16i\left[ \rlap{/}{\bar{x}} \sigma^{\mu\nu} \partial_\nu \bar{S}^{\rm ren.}\right] .
\ee
The associated conformal central charge density is defined by the $x$ and $y$ symmetric part of the $z$ integral of the current
\be
Y_{\alpha\dot\alpha} = \frac{1}{16}\int_{-\infty}^{+\infty} dz Y^0_{\alpha\dot\alpha} =-\sigma^3_{\alpha\dot\alpha}~ t \zeta ,
\ee
independent of radiative corrections.  

On the other hand the second space-time moment conformal tensorial central charge has radiative corrections.  From equation (\ref{tencentcurrents}) the conformal central charge current is given by
\be
X^\mu_{\alpha\beta} = -16i \left[\rlap{/}{x} \bar\sigma^{\mu\nu} \rlap{/}{\bar{x}} \right]_{\alpha\beta} \partial_\nu S^{\rm ren.} .
\ee
Hence the associated conformal central charge density is defined by the $x$ and $y$ symmetric part of the $z$ integral of the current
\be
X_{\alpha\beta} = \frac{1}{16} \int_{-\infty}^{+\infty} dz X^0_{\alpha\beta} = -\sigma^3_{\alpha\beta}\left[ (t^2 +x^2 +y^2) \zeta - \int_{-\infty}^{+\infty} dz z^2 \partial_z S^{\rm ren.} \right].
\ee
The first term similarly receives no radiative corrections, however the second term involves the non-vanishing $z$-integral of $S^{\rm ren.}= -2(\frac{m^3}{\sqrt{\lambda}})~\tanh{m\sqrt{\lambda}z}~[2/3 -1/3~\frac{\gamma }{\cosh^2{m\sqrt{\lambda}z}}]$ which has dependence on the anomalous dimension $\gamma$ of the fields.  The conformal tensorial central charges $X_{\alpha\beta}$ and $\bar{X}_{\dot\alpha\dot\beta}$ receive radiative corrections.

To conclude, some possible extensions of the current results are indicated. 
It seems natural to extend this work to include vector superfields and to consider the tensorial central charge extension of superconformal algebras with more than minimal SUSY. These directions are particularly appealing since
$D=4$, $N=2$ supersymmetric QCD has been found to admit
domain walls \cite{N=2-Abelian-wall} \cite{N=2-NA-wall} 
as well as non-Abelian vortex-strings \cite{N=2-NA-vortex}  
and their junctions (see \cite{N=2-composite} for a review). 
The central charge extension of the $N=2$ (softly broken) 
superconformal algebra can be studied by constructing the supercurrent as was done in this paper.  Along the same lines, it is of interest to investigate the construction of BPS solitons in 
the superconformal case of a $U(N)$ gauge theory with $2N$ hypermultiplets in the fundamental representation.  In addition, mass deformed $N=4$ superconformal gauge theories such as the $N=1^*$ or $N=2^*$ theories can be studied
with the methods developed here, as they have been shown to include BPS domain walls 
\cite{Bachas:2000dx}, \cite{Ritz:2006ji} 
and vortex-strings \cite{Markov:2004mj}.
In this regard it should be noted that the domain wall tensorial central charge in the SUSY algebra of the $N=1^*$ theory has recently been 
discussed in \cite{Ritz:2006ji}.

\section*{Acknowledgment}
The work of TEC was supported in part by the U.S. Department of Energy under grant DE-FG02-91ER40681 (Task B), and the work of TtV was supported in part by a Cottrell Award from the Research Corporation.
MN would like to thank the theoretical physics group at Purdue University and TtV would like to thank the theoretical physics groups at Purdue University and the Tokyo Institute of Technology for their hospitality during visits while this work was being completed.

\bigskip

\setcounter{newapp}{1}
\setcounter{equation}{0}
\renewcommand{\theequation}{\thenewapp.\arabic{equation}}

\section*{\large Appendix A: \, Tensorial Central Charge Extension\\
\hspace*{1.6in} of the Superconformal Algebra}

The tensorial central charge extension of the $N=1$, $D=4$ superconformal algebra consists of the charges
\begin{itemize}
\item $P^\mu$ : space-time translations
\item $M^{\mu\nu}$ : Lorentz transformations
\item $D$ : Dilatations
\item $K^\mu$ : (Special) Conformal transformations
\item $Q_\alpha$, ${\bar{Q}}_{\dot\alpha}$ : Supersymmetry transformations
\item $S_\alpha$, ${\bar{S}}_{\dot\alpha}$ : Special or Conformal Supersymmetry transformations
\item $R$ : $R$-transformations
\item $Z_{\alpha\beta}$, ${\bar{Z}}_{\dot\alpha\dot\beta}$ : SUSY Tensorial Central Charges
\item $X_{\alpha\beta}$, ${\bar{X}}_{\dot\alpha\dot\beta}$,
$Y_{\alpha\dot\alpha}$, ${\bar{Y}}_{\alpha\dot\alpha}$ : Conformal Tensorial Central Charges
\end{itemize}

The conformal algebra is given in terms of the 15 charges $P^\mu$, $M^{\mu\nu}$, $D$, and $K^\mu$:
\bea
[P^\mu , P^\nu ] &=& 0 \cr
[M^{\mu\nu}, M^{\rho\sigma}] &=& -i\left( g^{\mu\rho}M^{\nu\sigma} - g^{\mu\sigma}M^{\nu\rho} + g^{\nu\sigma}M^{\mu\rho} - g^{\nu\rho}M^{\mu\sigma}
\right) \cr
[M^{\mu\nu}, P^\lambda ] &=& i \left(  P^\mu g^{\nu\lambda} - P^\nu g^{\mu\lambda} \right) \cr
[M^{\mu\nu}, K^\lambda ] &=& i \left(  K^\mu g^{\nu\lambda} - K^\nu g^{\mu\lambda} \right) \cr
[M^{\mu\nu} , D ] &=& 0 \cr
[D, P^\mu] &=& -i P^\mu \qquad ; \qquad [D, K^\mu] = +i K^\mu  \cr
[P^\mu , K^\nu ] &=& +2i \left( g^{\mu\nu} D - M^{\mu\nu} \right) \cr
[K^\mu , K^\nu ] &=& 0 ,
\eea
where the Minkowski metric $g_{\mu\nu}$ has signature $(+,-,-,-)$.

The tensorial central charge extended super-Poincar\'e algebra includes the generators $Q_\alpha$, ${\bar{Q}}_{\dot\alpha}$, and $R$ and the SUSY tensorial central charges 
$Z_{\alpha\beta}$ and ${\bar{Z}}_{\dot\alpha\dot\beta}$:
\bea
\{ Q_\alpha , {\bar{Q}}_{\dot\alpha}\}&=& +2\sigma^\mu_{\alpha\dot\alpha}P_\mu \cr
[M^{\mu\nu}, Q_\alpha ] &=& -\frac{1}{2}\sigma^{\mu\nu~\beta}_\alpha Q_\beta \cr
[M^{\mu\nu}, \bar{Q}_{\dot\alpha} ] &=& -\frac{1}{2}\bar\sigma^{\mu\nu}_{\dot\alpha\dot\beta} \bar{Q}^{\dot\beta} \cr
[P^\mu , Q_\alpha ] &=& 0 = [P^\mu , \bar{Q}_{\dot\alpha} ] \cr
[R , P^\mu ] &=& 0 = [R, M^{\mu\nu} ] \cr
[R, Q_\alpha ] &=& +Q_\alpha \cr
[R, \bar{Q}_{\dot\alpha} ] &=& -\bar{Q}_{\dot\alpha}
\eea
and the central extension
\be
\{Q_\alpha , Q_\beta \} = Z_{\alpha\beta} \qquad ; \qquad \{\bar{Q}_{\dot\alpha} , \bar{Q}_{\dot\beta} \} = \bar{Z}_{\dot\alpha\dot\beta}  .
\ee
The graded Jacobi identity applied to $(M, Q, Q)$ implies that $Z$ is in the (1,0) representation of the Lorentz group while the Jacobi identity applied to $(M, \bar{Q}, \bar{Q})$ implies that $\bar{Z}$ transforms as the (0,1) representation of the Lorentz group.  Likewise, the graded Jacobi identity applied to $(R, Q, Q)$ and $(R, \bar{Q}, \bar{Q})$ yields that the $R$-weights of the charges are additive, hence 
\bea
[R, Z_{\alpha\beta} ] &=& +2 Z_{\alpha\beta} \cr
[R, \bar{Z}_{\dot\alpha\dot\beta} ] &=& -2 \bar{Z}_{\dot\alpha\dot\beta}.
\eea
Considering the Jacobi identities for $(P,Q,Q)$ and $(P,\bar{Q},\bar{Q})$ yields the translation invariance of $Z$ and $\bar{Z}$
\be
[P^\mu , Z_{\alpha\beta} ] = 0 = [P^\mu , \bar{Z}_{\dot\alpha\dot\beta} ] .
\ee

Including the dilatation, $D$, and conformal, $K^\mu$, transformations requires the introduction of the special supersymmetry transformation generators $S_\alpha$ and $\bar{S}_{\dot\alpha}$.  The relevant commutators are
\bea
[D, Q_\alpha ] &=& -\frac{i}{2} Q_\alpha \cr
[D, \bar{Q}_{\dot\alpha}] &=& -\frac{i}{2} \bar{Q}_{\dot\alpha} \cr
[K^\mu , Q_\alpha ] &=& -i\sigma^\mu_{\alpha\dot\alpha} \bar{S}^{\dot\alpha} \cr
[K^\mu , \bar{Q}_{\dot\alpha} ] &=& -iS^\alpha \sigma^\mu_{\alpha\dot\alpha} .
\eea
The graded Jacobi identities imply that the scaling weights of the charges also are additive 
\bea
[D, Z_{\alpha\beta}] &=& -iZ_{\alpha\beta} \cr
[D, \bar{Z}_{\dot\alpha\dot\beta}] &=& -i\bar{Z}_{\dot\alpha\dot\beta} .
\eea

In order to include the central charges $Z_{\alpha\beta}$ and $\bar{Z}_{\dot\alpha\dot\beta}$ in the superconformal algebra additional tensorial central charges need to be introduced.  In addition the supersymmetry and special supersymmetry charges' anticommutation relations must be modified.  Including the superconformal charges $S_\alpha$ and $\bar{S}_{\dot\alpha}$, the superconformal algebra begins as usual
\bea
[M^{\mu\nu}, S_\alpha ] &=& -\frac{1}{2}\sigma^{\mu\nu~\beta}_\alpha S_\beta \cr
[M^{\mu\nu}, \bar{S}_{\dot\alpha} ] &=& -\frac{1}{2}\bar\sigma^{\mu\nu}_{\dot\alpha\dot\beta} \bar{S}^{\dot\beta} \cr
[P^\mu , S_\alpha ] &=& -i \sigma^\mu_{\alpha\dot\alpha} \bar{Q}^{\dot\alpha} \cr
[P^\mu , \bar{S}_{\dot\alpha} ] &=& -iQ^\alpha \sigma^\mu_{\alpha\dot\alpha} \cr
[R, S_\alpha ] &=& -S_\alpha \cr
[R, \bar{S}_{\dot\alpha} ] &=& +\bar{S}_{\dot\alpha} \cr
[D, S_\alpha ] &=& +\frac{i}{2} S_\alpha \cr
[D, \bar{S}_{\dot\alpha}] &=& +\frac{i}{2} \bar{S}_{\dot\alpha} \cr
\{S_\alpha , \bar{S}_{\dot\alpha} \} &=& +2\sigma^\mu_{\alpha\dot\alpha} K_\mu .
\eea
The supersymmetry charges' algebra however must be altered to include the parameter $\rho$ which has the self consistent values determined by the graded Jacobi identity.  $\rho =1$ in the absence of tensorial central charges and $\rho =-1/3$ when the algebra is extended to include additional central charges $X_{\alpha\beta}$, ${\bar{X}}_{\dot\alpha\dot\beta}$, $Y_{\alpha\dot\alpha}$ and $\bar{Y}_{\alpha\dot\alpha}$
\bea
\{Q_\alpha , S_\beta \} &=& +i \left( \sigma^{\mu\nu}_{\alpha\beta} M_{\mu\nu} +2i\epsilon_{\alpha\beta} D
-3\rho \epsilon_{\alpha\beta} R \right) \cr
\{\bar{Q}_{\dot\alpha} , \bar{S}_{\dot\beta} \} &=& -i \left( \bar\sigma^{\mu\nu}_{\dot\alpha\dot\beta} M_{\mu\nu} -2i\epsilon_{\dot\alpha\dot\beta} D -3\rho \epsilon_{\dot\alpha\dot\beta} R \right) \cr
\{Q_\alpha , \bar{S}_{\dot\alpha} \} &=& Y_{\alpha\dot\alpha} = +2\sigma^\mu_{\alpha\dot\alpha} Y_\mu \cr
\{S_\alpha , \bar{Q}_{\dot\alpha} \} &=& \bar{Y}_{\alpha\dot\alpha} = +2\sigma^\mu_{\alpha\dot\alpha} \bar{Y}_\mu \cr
\{S_\alpha , S_\beta \} &=& X_{\alpha\beta} \cr
\{\bar{S}_{\dot\alpha} , \bar{S}_{\dot\beta} \} &=& \bar{X}_{\dot\alpha\dot\beta}  .
\label{rhofactor}
\eea

As above, the Jacobi identities imply that $Y$ and $\bar{Y}$ are in the $(1/2 , 1/2)$ representation of the Lorentz group while $X$ and $\bar{X}$ are in the (1,0) and (0,1) representations.  The $R$-weights and scaling weights act additively, hence the following table of the various weights (including the non-central charges) is obtained.\linebreak

\begin{center}
\begin{tabular}{| c | c |c |}
\hline Charge & Scale Weight & $R$-Weight 
\raisebox{-2ex}{\rule{0cm}{5ex}}\\
 & (Dimension) &  
\raisebox{-2ex}{\rule{0cm}{5ex}}\\
\hline\hline $P^\mu$ & -1 & 0
\raisebox{-2ex}{\rule{0cm}{5ex}}\\
\hline $M^{\mu\nu}$ & 0 & 0
\raisebox{-2ex}{\rule{0cm}{5ex}}\\ 
\hline $D$ & 0 & 0
\raisebox{-2ex}{\rule{0cm}{5ex}}\\
\hline $K^\mu$ & +1 & 0
\raisebox{-2ex}{\rule{0cm}{5ex}}\\
\hline $R$ & 0 & 0
\raisebox{-2ex}{\rule{0cm}{5ex}}\\
\hline $Q_\alpha$ & -1/2 & +1 
\raisebox{-2ex}{\rule{0cm}{5ex}}\\
\hline $\bar{Q}_{\dot\alpha}$ & -1/2 & -1 
\raisebox{-2ex}{\rule{0cm}{5ex}}\\
\hline $S_\alpha$ & +1/2 & -1 
\raisebox{-2ex}{\rule{0cm}{5ex}}\\
\hline $\bar{S}_{\dot\alpha}$ & +1/2 & +1 
\raisebox{-2ex}{\rule{0cm}{5ex}}\\
\hline $Z_{\alpha\beta}$ & -1 & +2
\raisebox{-2ex}{\rule{0cm}{5ex}}\\
\hline $\bar{Z}_{\dot\alpha\dot\beta}$ & -1 & -2
\raisebox{-2ex}{\rule{0cm}{5ex}}\\
\hline $X_{\alpha\beta}$ & +1 & -2
\raisebox{-2ex}{\rule{0cm}{5ex}}\\
\hline $\bar{X}_{\dot\alpha\dot\beta}$ & +1 & +2
\raisebox{-2ex}{\rule{0cm}{5ex}}\\
\hline $Y_{\alpha\dot\alpha}$ & 0 & +2
\raisebox{-2ex}{\rule{0cm}{5ex}}\\
\hline $\bar{Y}_{\alpha\dot\alpha}$ & 0 & -2
\raisebox{-2ex}{\rule{0cm}{5ex}}\\
\hline
\end{tabular}
\end{center}
\bigskip

Additional use of the Jacobi identities determines the commutation relations involving the central charges as well as the consistency requirement for the parameter $\rho$.  For the tensorial central charge extension of the $N=1,~ D=4$ superconformal algebra to be consistent $\rho = -1/3$, without tensorial central charges $\rho = 1$ for consistency.  First, the non-trivial commutators involving the central charges are listed along with the Jacobi identities involved.  The restrictions of the scale weight of charges being $0,~\pm 1/2,~\pm 1$ and the $R$-weight of charges being $0,~\pm 1,~\pm 2$ set many commutators to zero.  First the space-time transformations of the central charges are considered.  This exhibits the necessity of introducing the set of tensorial central charges $\{Z, \bar{Z}, X, \bar{X}, Y, \bar{Y}\}$
\bea
(P,Q,\bar{S}):~ [ P^\mu, Y_{\alpha\dot\alpha} ] &=& +i \bar\sigma^{\mu~\beta}_{\dot\alpha} Z_{\beta\alpha} \cr
(P,\bar{Q},S):~ [ P^\mu, \bar{Y}_{\alpha\dot\alpha} ] &=& +i \sigma^{\mu~\dot\beta}_{\alpha} 
\bar{Z}_{\dot\alpha\dot\beta} \cr
(P,S,S):~ [P^\mu , X_{\alpha\beta} ] &=& -i\sigma^\mu_{\alpha\dot\alpha} \bar{Y}_\beta^{~\dot\alpha}
-i\sigma^\mu_{\beta\dot\alpha} \bar{Y}_\alpha^{~\dot\alpha} \cr
(P,\bar{S},\bar{S}):~ [P^\mu , \bar{X}_{\dot\alpha\dot\beta} ] &=& +i\bar\sigma^{\mu~\alpha}_{\dot\alpha} 
Y_{\alpha\dot\beta} +i\bar\sigma^{\mu~\alpha}_{\dot\beta} Y_{\alpha\dot\alpha} \cr
(K,Q,Q):~ [K^\mu , Z_{\alpha\beta}] &=& +iY_{\alpha\dot\beta} \bar\sigma^{\mu\dot\beta}_{~~\beta} +iY_{\beta\dot\beta} 
\bar\sigma^{\mu\dot\beta}_{~~\alpha}  \cr
(K,\bar{Q},\bar{Q}):~ [K^\mu , \bar{Z}_{\dot\alpha\dot\beta}] &=& +i \bar\sigma^{\mu\alpha}_{\dot\beta}
\bar{Y}_{\alpha\dot\alpha} +i\bar\sigma^{\mu\alpha}_{\dot\alpha} \bar{Y}_{\alpha\dot\beta}  \cr
(K,Q,\bar{S}):~ [K^\mu , Y_{\alpha\dot\alpha} ] &=& +i \bar{X}_{\dot\alpha\dot\beta} \sigma^{\mu\dot\beta}_{\alpha} \cr
(K,\bar{Q},S):~ [K^\mu , \bar{Y}_{\alpha\dot\alpha} ] &=& +i \bar\sigma^{\mu\beta}_{\dot\alpha} {X}_{\beta\alpha}
\eea

The supersymmetry and superconformal transformations of the central charges are found to be
\bea
(Q,\bar{Q},\bar{S}):~ [\bar{Q}_{\dot\alpha} , Y_{\beta\dot\beta} ] &=& -3i(1-\rho) \epsilon_{\dot\alpha\dot\beta} Q_\beta \cr
(Q,\bar{Q},S):~ [{Q}_{\alpha} , \bar{Y}_{\beta\dot\beta} ] &=& -3i(1-\rho) \epsilon_{\alpha\beta} \bar{Q}_{\dot\beta} \cr
(Q,S,S):~ [Q_\gamma , X_{\alpha\beta} ] &=& -3i (1-\rho) \left(\epsilon_{\gamma\alpha} S_{\beta} +\epsilon_{\gamma\beta} S_{\alpha} \right)  \cr
(\bar{Q},\bar{S},\bar{S}):~ [\bar{Q}_{\dot\gamma} , \bar{X}_{\dot\alpha\dot\beta} ] &=& -3i (1-\rho) \left(
\epsilon_{\dot\gamma\dot\alpha} \bar{S}_{\dot\beta} +\epsilon_{\dot\gamma\dot\beta} \bar{S}_{\dot\alpha} \right)  \cr
(Q,Q,S):~ [ S_\gamma , Z_{\alpha\beta} ] &=& -3i (1-\rho) \left(\epsilon_{\gamma\alpha} Q_{\beta} +\epsilon_{\gamma\beta} Q_{\alpha} \right)  \cr
(\bar{Q},\bar{Q},\bar{S}):~ [\bar{S}_{\dot\gamma} , \bar{Z}_{\dot\alpha\dot\beta} ] &=& -3i (1-\rho) \left(
\epsilon_{\dot\gamma\dot\alpha} \bar{Q}_{\dot\beta} +\epsilon_{\dot\gamma\dot\beta} \bar{Q}_{\dot\alpha} \right)  \cr
(Q,S,\bar{S}):~ [{S}_{\alpha} , Y_{\beta\dot\beta} ] &=& -3i(1-\rho) \epsilon_{\alpha\beta} \bar{S}_{\dot\beta} \cr
(\bar{Q},S,\bar{S}):~ [\bar{S}_{\dot\alpha} , \bar{Y}_{\beta\dot\beta} ] &=& -3i(1-\rho) \epsilon_{\dot\alpha\dot\beta} {S}_{\beta} .
\label{rhoeq1}
\eea

Finally, the commutators of the central charges with each other
\bea
(Q,Q,\bar{Y}),~(\bar{Q},S,Z) :~ [Z_{\alpha\beta} , \bar{Y}_{\gamma\dot\gamma} ] &=& -6i (1-\rho) \left( \epsilon_{\alpha\gamma} \sigma^\mu_{\beta\dot\gamma} +\epsilon_{\beta\gamma} \sigma^\mu_{\alpha\dot\gamma} \right) P_\mu  \cr
(\bar{Q},\bar{Q},{Y}), ~({Q},\bar{S},\bar{Z}):~ [\bar{Z}_{\dot\alpha\dot\beta} , {Y}_{\gamma\dot\gamma} ] &=& -6i (1-\rho) \left( 
\epsilon_{\dot\alpha\dot\gamma} \sigma^\mu_{\gamma\dot\beta} +\epsilon_{\dot\beta\dot\gamma} 
\sigma^\mu_{\gamma\dot\alpha} \right) P_\mu  \cr
(S,S,Y),~(Q,\bar{S},X):~ [X_{\alpha\beta} , Y_{\gamma\dot\gamma} ] &=& -6i (1-\rho) \left( \epsilon_{\alpha\gamma} 
\sigma^\mu_{\beta\dot\gamma} +\epsilon_{\beta\gamma} \sigma^\mu_{\alpha\dot\gamma} \right) K_\mu  \cr
(\bar{Q},S,\bar{X}),~(\bar{S},\bar{S},\bar{Y}):~ [\bar{X}_{\dot\alpha\dot\beta} , \bar{Y}_{\gamma\dot\gamma} ] &=& -6i (1-\rho) \left( 
\epsilon_{\dot\alpha\dot\gamma} \sigma^\mu_{\gamma\dot\beta} +\epsilon_{\dot\beta\dot\gamma} 
\sigma^\mu_{\gamma\dot\alpha} \right) K_\mu  \cr
(Q,\bar{S},\bar{Y}),~(\bar{Q},S,Y):~ [ Y_{\alpha\dot\alpha} , \bar{Y}_{\beta\dot\beta} ] &=& -3(1-\rho) \left( \epsilon_{\alpha\beta}
\bar\sigma^{\mu\nu}_{\dot\beta\dot\alpha} +\sigma^{\mu\nu}_{\alpha\beta} \epsilon_{\dot\beta\dot\alpha} \right) 
M_{\mu\nu}  \cr
 & &\qquad\qquad +18\rho (1-\rho ) \epsilon_{\alpha\beta} \epsilon_{\dot\beta\dot\alpha} R  \cr
(Q,Q,X),~(S,S,Z):~ [Z_{\alpha\beta}, X_{\gamma\delta} ] &=& +3(1-\rho ) M_{\mu\nu} \left( 
\epsilon_{\alpha\gamma} \sigma^{\mu\nu}_{\beta\delta} 
+ \epsilon_{\beta\gamma} \sigma^{\mu\nu}_{\alpha\delta} 
+ \epsilon_{\alpha\delta} \sigma^{\mu\nu}_{\beta\gamma} 
+ \epsilon_{\beta\delta} \sigma^{\mu\nu}_{\alpha\gamma} \right)  \cr
 & & \qquad\qquad +12i (1-\rho ) D \left( \epsilon_{\alpha\gamma} \epsilon_{\beta\delta} +
\epsilon_{\alpha\delta} \epsilon_{\beta\gamma} \right) \cr
  & & \qquad\qquad\qquad  -18\rho (1-\rho ) R \left( \epsilon_{\alpha\gamma} \epsilon_{\beta\delta} +
\epsilon_{\alpha\delta} \epsilon_{\beta\gamma} \right)  \cr
(\bar{S},\bar{S},\bar{Z}),~(\bar{Q},\bar{Q},\bar{X}):~ [\bar{Z}_{\dot\alpha\dot\beta}, \bar{X}_{\dot\gamma\dot\delta} ] &=& -3(1-\rho ) M_{\mu\nu} \left( 
\epsilon_{\dot\alpha\dot\gamma} \bar\sigma^{\mu\nu}_{\dot\beta\dot\delta} 
+ \epsilon_{\dot\beta\dot\gamma} \bar\sigma^{\mu\nu}_{\dot\alpha\dot\delta} 
+ \epsilon_{\dot\alpha\dot\delta} \bar\sigma^{\mu\nu}_{\dot\beta\dot\gamma} 
+ \epsilon_{\dot\beta\dot\delta} \bar\sigma^{\mu\nu}_{\dot\alpha\dot\gamma} \right)  \cr
  & & \qquad\qquad +12i (1-\rho )D \left( \epsilon_{\dot\alpha\dot\gamma} \epsilon_{\dot\beta\dot\delta} +
\epsilon_{\dot\alpha\dot\delta} \epsilon_{\dot\beta\dot\gamma} \right) \cr
  & & \qquad\qquad\qquad  +18\rho (1-\rho ) R \left( \epsilon_{\dot\alpha\dot\gamma} \epsilon_{\dot\beta\dot\delta} +
\epsilon_{\dot\alpha\dot\delta} \epsilon_{\dot\beta\dot\gamma} \right)  .\cr
 & & 
\eea

The remaining Jacobi identities provide an important consistency check for the above algebra.  Indeed, each of the remaining Jacobi triplets that involve SUSY charge, conformal SUSY charge and at least one tensorial central charge, for example $(Q, S, \bar{Z})$ or $(Q, \bar{X}, \bar{Y})$,  yields 
\be
\rho = -\frac{1}{3} 
\ee
in the case of non-trivial tensorial central charges or they provide a consistency check independent of the choice of $\rho (\neq 1)$.  In the absence of tensorial central charges, the Jacobi closure of the algebra requires, as usual, that $\rho =1$ as seen in equation (\ref{rhoeq1}).  This completes the tensorial central charge extension of the $N=1, ~D=4$ superconformal algebra, 
$OSp(1|8)$ \cite{Ferrara:1999si}, \cite{vanHolten:1982mx}, \cite{Bedding:1983uf}.


\setcounter{newapp}{2}
\setcounter{equation}{0}
\renewcommand{\theequation}{\thenewapp.\arabic{equation}}

\section*{\large\bf Appendix B: \, Superconformal Transformations\\
 \hspace*{2.5in} and Component Fields}

Of particular interest are the SUSY and superconformal transformations of the superfields and component fields.  The SUSY transformations of the superfields are represented by superspace differential operators acting on the fields (suppressing the internal indices)
\bea
i[Q_\alpha , \phi] &=& \delta^Q_\alpha \phi = \left( \frac{\partial}{\partial \theta^\alpha} + i(\rlap{/}{\partial}\bar\theta)_\alpha \right) \phi \cr
i[\bar{Q}_{\dot\alpha} , \phi] &=& \delta^{\bar{Q}}_{\dot\alpha} \phi = \left( -\frac{\partial}{\partial \bar\theta^{\dot\alpha}} - i(\theta\rlap{/}{\partial})_{\dot\alpha} \right) \phi ,
\eea
and likewise for $\bar\phi$.  Expanding in terms of components yields the usual SUSY variation of the component fields
\bea
\delta^{Q}_{\alpha} A &=& \psi_{\alpha} ~\qquad\qquad\qquad \delta^{Q}_{\alpha} \bar{A} =0\cr
\delta^{Q}_{\alpha} \psi_{\beta} &=& -2 \epsilon_{\alpha\beta} F \qquad\qquad \delta^{Q}_{\alpha} \bar\psi_{\dot\beta} = +2i {\rlap{/}{\partial}}_{\alpha\dot\beta} \bar{A} \cr
\delta^{Q}_{\alpha} F &=& 0 \qquad\qquad\quad\qquad \delta^{Q}_{\alpha} \bar{F} = -i(\rlap{/}{\partial}\bar\psi)_{\alpha} \cr
 & & \cr
\delta^{\bar{Q}}_{\dot\alpha} A &=& 0 ~\qquad\qquad\qquad\quad \delta^{\bar{Q}}_{\dot\alpha} \bar{A} =\bar\psi_{\dot\alpha}\cr
\delta^{\bar{Q}}_{\dot\alpha} \psi_\alpha &=& +2i{\rlap{/}{\partial}}_{\alpha\dot\alpha}A \qquad\qquad \delta^{\bar{Q}}_{\dot\alpha} \bar{\psi}_{\dot\beta} =+2 \epsilon_{\dot\alpha\dot\beta}\bar{F}\cr
\delta^{\bar{Q}}_{\dot\alpha} F &=& +i(\partial_\rho \psi \sigma^\rho)_{\dot\alpha}  \quad\qquad \delta^{\bar{Q}}_{\dot\alpha} \bar{F} = 0 .
\eea

The superconformal transformations of the fields can be represented by the superspace differential operators
\bea
i[S_\alpha , \phi ] &=& \delta^S_\alpha \phi = \left( -x_\mu \sigma^\mu \delta^{\bar{Q}} +2\theta \delta^R -i\theta^2 D -2i(d -\frac{n}{2})\theta \right)_\alpha \phi \cr
i[\bar{S}_{\dot\alpha} , \phi ] &=& \delta^{\bar{S}}_{\dot\alpha} \phi = \left( -x_\mu \delta^Q \sigma^\mu  +2\bar\theta \delta^R +i\bar\theta^2 \bar{D} +2i(d +\frac{n}{2})\bar\theta \right)_{\dot\alpha} \phi ,
\eea
where the $R$ transformation for superfields is
\be
i[ R , \phi ] = \delta^R \phi = i\left( n + \theta \frac{\partial}{\partial \theta} + \bar\theta \frac{\partial}{\partial \bar\theta} \right) \phi ~.
\ee
Chirality of the fields requires the $R$ weight $n$ to be related to the scale weight $d$ of the fields: $n= -\frac{2}{3}d$ for chiral superfields and $\bar{n}=+\frac{2}{3}d$ for antichiral superfields.  The $R$ transformations for the component fields become
\bea
\delta^R A &=& in A \qquad\qquad\qquad\qquad~ \delta^R \bar{A} = -i \bar{n}\bar{A} \cr
\delta^R \psi &=& i(n+1) \psi \qquad\qquad\qquad \delta^R \bar{\psi} = -i(\bar{n}+1)\bar{\psi} \cr
\delta^R F &=& i(n+2) F \qquad\qquad\qquad \delta^R \bar{F} = -i(\bar{n}+2)\bar{F} .
\eea
Likewise, expanding in terms of component fields, the superconformal variations of the component fields are secured
\bea
\delta^{S}_{\alpha} A &=& (-\rlap{/}{x} \delta^{\bar{Q}})_\alpha A = 0\cr
\delta^{S}_{\alpha} \psi_{\beta} &=& (-\rlap{/}{x} \delta^{\bar{Q}})_\alpha \psi_\beta -4id \epsilon_{\alpha\beta} A= 2i (\rlap{/}{x} \bar{\rlap{/}{\partial}}A)_{\alpha\beta}-4id \epsilon_{\alpha\beta} A \cr
\delta^{S}_{\alpha} F &=& (-\rlap{/}{x} \delta^{\bar{Q}})_\alpha F -2i(1-d)\psi_\alpha= -i(\rlap{/}{x} \bar{\rlap{/}{\partial}}\psi)_{\alpha} -2i(1-d)\psi_\alpha \cr 
 & & \cr
\delta^{S}_{\alpha} \bar{A} &=&(-\rlap{/}{x} \delta^{\bar{Q}})_\alpha A = -(\rlap{/}{x}\bar\psi)\cr
\delta^{S}_{\alpha} \bar\psi_{\dot\beta} &=& (-\rlap{/}{x} \delta^{\bar{Q}})_\alpha \bar\psi_{\dot\beta} = 2\rlap{/}{x}_{\alpha\dot\beta}\bar{F}\cr
\delta^{S}_{\alpha} \bar{F} &=& (-\rlap{/}{x} \delta^{\bar{Q}})_\alpha \bar{F}=0 \cr
 & & \cr
 & & \cr
\delta^{\bar{S}}_{\dot\alpha} A &=& (\bar{\rlap{/}{x}}\delta^Q)_{\dot\alpha} A = (\bar{\rlap{/}{x}}\psi)_{\dot\alpha}\cr
\delta^{\bar{S}}_{\dot\alpha} \psi_\alpha &=& (\bar{\rlap{/}{x}}\delta^Q)_{\dot\alpha}\psi_\alpha = 2{\rlap{/}{x}}_{\alpha\dot\alpha} F\cr 
\delta^{\bar{S}}_{\dot\alpha} F &=& (\bar{\rlap{/}{x}}\delta^Q)_{\dot\alpha} F =0  \cr 
 & & \cr
\delta^{\bar{S}}_{\dot\alpha} \bar{A} &=&(\bar{\rlap{/}{x}}\delta^Q)_{\dot\alpha}\bar{A} =0\cr
\delta^{\bar{S}}_{\dot\alpha} \bar{\psi}_{\dot\beta} &=&(\bar{\rlap{/}{x}}\delta^Q)_{\dot\alpha}\bar\psi_{\dot\beta} +4id \epsilon_{\dot\alpha\dot\beta} \bar{A} = 2i(\bar{\rlap{/}{x}}\rlap{/}{\partial} \bar{A} )_{\dot\alpha\dot\beta} +4id \epsilon_{\dot\alpha\dot\beta} \bar{A} \cr
\delta^{\bar{S}}_{\dot\alpha} \bar{F} &=& (\bar{\rlap{/}{x}}\delta^Q)_{\dot\alpha}\bar{F} + 2i (1-d)\bar\psi_{\dot\alpha}= -i( \bar{\rlap{/}{x}} \rlap{/}{\partial}\bar\psi)_{\dot\alpha} + 2i (1-d)\bar\psi_{\dot\alpha}.
\eea

The spacetime translation symmetry transformations have the universal form
\be
i[ P_\mu , \phi ] = \delta^{P}_{\mu}\phi = \partial_\mu \phi 
\ee
and likewise for the antichiral field.  Hence each component field transforms as the spacetime derivative
\be
\delta^P_\mu \varphi = \partial_\mu  \varphi ,
\ee
where $\varphi$ runs over all component fields, $\varphi \in \{A, \psi, F, \bar{A}, \bar\psi, \bar{F}\}$.

The generalized Wess-Zumino superspace action, equation (\ref{WZModel1}),
\be
\Gamma = \int dV ~K(\phi , \bar\phi) + \int dS ~W(\phi) + \int d\bar{S} ~\bar{W}(\bar\phi),
\ee
has the component field, equation (\ref{compfields}), expansion
\be
\Gamma = \int d^4 x {\cal L},
\label{componentaction}
\ee
with Lagrangian 
\bea
{\cal L} &=& 16 K,_{i\bar{i}} \partial_\mu A^i \partial^\mu \bar{A}^{\bar{i}} + 16 K,_{i\bar{i}} F^i \bar{F}^{\bar{i}} + 4i K,_{i\bar{i}} (\psi^i \stackrel{\leftrightarrow}{\rlap{/}{\partial}}\bar\psi^{\bar{i}} ) \cr
 & & \cr
 & &+\psi^i \psi^j K,_{ij\bar{i}\bar{j}} \bar\psi^{\bar{i}} \bar\psi^{\bar{j}} - 4 \psi^i \psi^j K,_{ij\bar{i}} \bar{F}^{\bar{i}} -4 F^i K,_{i\bar{i}\bar{j}} \bar\psi^{\bar{i}} \bar\psi^{\bar{j}}  \cr
 & & \cr
 & &-4i K,_{ij\bar{i}} \psi^i (\rlap{/}{\partial} A^j) \bar\psi^{\bar{i}} + 4i K,_{i\bar{i}\bar{j}} \psi^i ( \rlap{/}{\partial} \bar{A}^{\bar{j}}) \bar\psi^{\bar{i}} \cr
 & & \cr
 & &-4 F^i W,_i -4 \bar{F}^{\bar{i}}\bar{W},_{\bar{i}} + W,_{ij} \psi^i \psi^j + \bar{W},_{\bar{i}\bar{j}} \bar\psi^{\bar{i}} \bar\psi^{\bar{j}} .
\label{componentlag}
\eea

\bigskip


\newpage
\end{document}